\documentclass[a4paper,english,aps,pra,superscriptaddress,twocolumn,accepted=2022-01-11,a4paper]{quantumarticle}
\pdfoutput=1

\usepackage[T1]{fontenc}
\usepackage[latin9]{inputenc}
\usepackage{units}
\usepackage{amssymb}
\usepackage{amsmath}
\usepackage{braket}
\usepackage{bbold}
\usepackage{graphicx}
\usepackage{color}
\usepackage{natbib}
\usepackage{babel}
\usepackage{array}
\usepackage{multirow}
\usepackage{placeins}
\usepackage{tikz}

\usepackage[ruled,vlined,noline]{algorithm2e}

\usepackage[
colorlinks=true,
linkcolor=darkgray,
urlcolor=darkgray,
citecolor=blue]{hyperref}

\pdfoutput=1
\graphicspath{./figures/}

\newcommand{\Rx}{X}
\newcommand{\Ry}{Y}
\newcommand{\Rz}{Z}

\newcommand{\instate}{\ket{1}\ket{+}^{\otimes 6}}

\usepackage[titles]{tocloft}
\cftsetindents{section}{0em}{2.5em}
\cftsetindents{subsection}{1em}{2em}

\usepackage{soul}
\newcommand{\change}[1]{\hl{#1}}
\renewcommand{\change}[1]{#1}

\begin{document}

\title{\change{Robust} quantum compilation and circuit optimisation via energy \change{minimisation}}

\author{Tyson Jones}
\affiliation{Department of Materials, University of Oxford, Parks Road, Oxford OX1 3PH, UK}
\email{tyson.jones@materials.ox.ac.uk}
\orcid{0000-0002-9360-5417}
	
\author{Simon C. Benjamin}
\email{simon.benjamin@materials.ox.ac.uk}
\orcid{0000-0002-7766-5348}
\affiliation{Department of Materials, University of Oxford, Parks Road, Oxford OX1 3PH, UK}

\begin{abstract}
We explore a method for automatically recompiling a quantum circuit $\mathcal{A}$ into a target circuit $\mathcal{B}$, with the goal that both circuits have the same action on a specific input i.e. $\mathcal{B}\ket{\text{in}}=\mathcal{A}\ket{\text{in}}$. This is of particular relevance to hybrid, NISQ-era algorithms for dynamical simulation or eigensolving.  
The user initially specifies  $\mathcal{B}$ as a blank template: a layout of parameterised unitary gates configured to the identity. 
The compilation then proceeds using quantum hardware to perform an isomorphic energy-minimisation task, and an optional gate elimination phase to compress the circuit. If $\mathcal{B}$ is insufficient for perfect recompilation then the method will result in an approximate solution. \change{
We optimise using imaginary time evolution, and a recent extension of quantum natural gradient for noisy settings.
We successfully recompile a $7$-qubit circuit involving $186$ gates of multiple types into an alternative form with a different topology, far fewer two-qubit gates, and a smaller family of gate types. Moreover we verify that the process is \textit{robust}, finding that per-gate noise of up to $1\%$ can still yield near-perfect recompilation. }
We test the scaling of our algorithm on up to $20$ qubits, recompiling into circuits with up to $400$ parameterized gates, and incorporate a \change{custom} adaptive timestep technique.
We note that a classical simulation of the process can be useful to optimise circuits for today's prototypes, and more generally the method may enable `blind' compilation i.e. harnessing a device whose response to control parameters is deterministic but unknown.
\vspace{.15cm} \\
The code and resources used to generate our results are openly available \href{https://github.com/QTechTheory/DissipativeRecompiler}{online}~\cite{githubLink,mmaGithubLink}. A simple Mathematica demonstration of our algorithm can be found at \href{https://questlink.qtechtheory.org}{questlink.qtechtheory.org}.
\end{abstract}
\maketitle

\pagebreak
\tableofcontents

\section{Introduction}

In conventional computing, compilers are essential to translate programs into efficient low-level instructions for execution at the hardware level. 
Quantum computers will also benefit greatly from efficient compilation, but the nature of the compilation goal depends on the nature of the machine. In the immediate future we are entering the so-called NISQ era of noisy, intermediate scale quantum devices~{\cite{preskill_nisq}}. In the further future we expect to see the emergence of fully fault-tolerant, code-protected quantum processors which can be regarded by programmers as error-free.

For fault-tolerant devices, typically only a limited non-universal family of operations can be performed directly on encoded data, so that other (usually non-Clifford) operations must be performed with the use of additional resources such as magic states~\cite{Bravyi2005,Li2015,Cambell2017}. Therefore the priority for computation will be to minimise the number of these expensive resource-consuming gates. Substantial efforts have been made to understand how to minimise the number of non-Clifford gates, such as $T-$gates, that are required to perform a given task~\cite{Gosset2014,Ross2016,Amy2016,Heyfron2017,Nam2018}. 

For the era of Noisy Intermediate Scale Quantum (NISQ) devices, the priorities for compilation will be different. Here we may expect that information is stored without the full protection of error-correcting codes and that therefore the difficulty that codes cannot permit universal operations does not arise. 
Rather, the costly gates are those with the greatest error burden. Typically these are two-qubit gates (and higher degree gates) in today's prototypes, while single-qubit gates are higher fidelity~\cite{Barends2014,Harty2014,Lucas,Wineland}. Moreover a given physical device will have certain operations that are native to it, so that e.g. it may be that a control-{\small NOT} is impossible to implement directly but is instead realised though a parity-dependent phase shift together with additional single-qubit gates. Furthermore the device will have a native connectivity: certain qubits will be able to directly link to others, for example in a two-dimensional nearest-neighbour topology or a more flexible networked architecture~\cite{Nickerson2014}. Thus one would wish to compile directly to the device's native gate set and connectivity. 

To take further the remark about a device's native operations: it may be that issues such as cross-talk mean that an operation targeted to a specific qubit inevitably leads to an unwanted, but deterministic, effect on proximal qubits. One can regard that operation as a kind of native gate itself (see e.g. Ref.~\cite{Heya2018}), albeit one that may be difficult to work with in analytic treatment. A compiler that is capable of targeting an arbitrary family of gates could recompile from a standard `white board' description of a circuit into a truly native format where the gate operations are bespoke for a specific device. As we presently discuss, taking this idea further one could compile into a device whose gates are an unknown function of the control parameters. 

This paper describes a general method of translating one quantum circuit into another, i.e. recompiling it. \change{The approach can in principle allow one to do the following}:
\begin{itemize}
\item Target an arbitrary (user-specified) circuit layout,
\item Target an arbitrary (user-specified) set of gates, including bespoke gates not used in analytic treatments,
\item Support approximate recompilation, so that if the specified target template is too shallow for perfect recompilation then an approximate circuit will be found,
\item Minimise the impact of noise.
\end{itemize}
We note that if one wished, the user-specification of the target layout and gates could in future be relaxed following VQA studies using ansatze that auto-evolve ~\cite{ADAPT,eve,vanAlg}.

The present scheme is limited in important ways:
\begin{itemize}
\item Compilation of circuits beyond the classical simulation limit will require quantum hardware, and will consume considerable time on that hardware. We note that all-classical software to recompile circuits involving parameterised gates does exist~\cite{Nam2018} and can make significant savings. However no classical compiler can be expected to approach optimality for general circuits since even the task of verifying that two circuits are near-identical is QMA-complete~\cite{Janzing2003}. 
\item Compilation from the original circuit $\mathcal{A}$ is not to an equivalent unitary circuit, but rather to a target circuit $\mathcal{B}$ that (ideally) has the same effect on just one specific input state $\ket{\text{in}}$, so that $\mathcal{B}\ket{\text{in}}=\mathcal{A}\ket{\text{in}}$.\\
This is a profoundly more permissive goal, but is in fact the {\em right} goal for many quantum algorithms including so-called hybrid quantum-classical approaches~\cite{chemReview2018}.
\item While the specific input state $\ket{\text{in}}$ can have any form, it is necessary that we `understand' it well enough to be able to write down a (fictitious) Hamiltonian for which it is the ground state. 
\item The approach we describe here has unproven scaling as the circuits grow beyond $20$ qubits. We remark on this point presently.
\item We restrict our attention to circuits formed of unitary gates, so that our complete circuits $\mathcal{A}$ and  $\mathcal{B}$ are themselves unitary. Generalising to non-unitary circuits would appear possible however.
\item A more comprehensive compiler might automatically propose and test different templates for $\mathcal{B}$, rather than requiring the user to specify one. This would be a higher-level process operating above the compilation we describe; prior work on all-classical optimisation could be employed here~\cite{Nam2018}.
\end{itemize}

We note other recent efforts to perform approximate recompilation of quantum circuits.
For example, a scheme presented by Khatri \textit{et al} describes a method to recompile a (possibly unknown) $n$-qubit unitary $U$ into a parametrised circuit $V$ {\cite{overlap_recomp}}, by maximising their \textit{overlap}. Their cost function is $|\text{Tr}(V^\dagger U)|^2$, evaluated via the Hilbert-Schmidt test, requiring $2n$ qubits and $2n$ additional controlled gates. They demonstrate their technique with noise-free and noisy simulations of up to $9$ qubits.
\change{They thus focus on the less permissive task of recompiling the entire unitary as opposed to its action on a fixed input state, while noting the possibility of fixed-input schemes.  Compared to the present work, that study}~\cite{overlap_recomp} \change{presents a broader range of techniques and includes 1-qubit demonstrations on cloud-based quantum hardware, but investigates considerably more shallow circuits as the input (e.g. $4$ rather than $60$ layers) and smaller registers (simulating a maximum of $9$ qubits rather than the $20$ considered here). To perform our larger and more complex compilation tasks, we introduce certain enabling techniques described presently, including adaptive timesteps, elimination of redundant gates, and a `lure' method to improve convergence. }


Meanwhile an experimentally-focused effort~{\cite{experi_recomp}} performed a similar scheme on a photonic processor of $4$ qubits, with numerical testing on up to $5$ qubits. There, they recompile an unknown unitary $U$ acting on a fixed, restrictedly seperable state $\ket{\text{in}}$, and use an ansatz divided into multiple layers $V_l$, each targeting one fewer qubits. The parameters in each layer are successively trained by driving the layer's nominated qubit toward its initial state in the input separable state, i.e. toward parameters which satisfy $(\prod_l V_l)^\dagger U \ket{\text{in}} \approx \ket{\text{in}}$.


In contrast to these works, our fixed-input method requires only $n$ qubits, with a cheap cost function evaluated via measuring a fictitious Hamiltonian. We make no conditions on the input state. When our input unitary happens to be parameterized, our scheme is compatible with an additional novel \textit{luring} technique, which can significantly improve convergence on systems larger than tested by the previous works. We test our scheme on up to $20$ qubits, and perform a demonstration of a possible application; extending variational real-time simulation. \change{We also make a comprehensive, density matrix based analysis of the impact of gate noise for a scenario involving over $200$ gates.}

\section{Variational Simulation}
\label{sec:vari_sim_overview}

\change{ 
Our recompilation method is an instance of a variational quantum algorithm; VQA's have been explored widely for their compatibility with first generation quantum computers. Two recent reviews of VQAs are available which cover the extensive and growing literature}~\cite{natPhyRevVQA,hybridVQAandMiti}.
\change{
As a subroutine, our algorithm performs energy minimisation, for which there are many quantum algorithms}~{\cite{Yuan2019,chemReview2018}}
\change{, though in this work we opt to use deterministic imaginary-time evolution~{\cite{imag2018}} and a noise-compatible form~{\cite{noisy_natural}} of quantum natural gradient~{\cite{natural_gradient}}, which we introduce below.
To demonstrate an interesting application of our recompilation strategy (Section~{\ref{sec:numerical_demo}}), we will employ a variational real-time simulation algorithm to produce input states. We also summarise this algorithm, which we refer to as Li's method~{\cite{YingPRX}}.
}

\subsection{Li's real time simulation}

\change{
Li's method~{\cite{YingPRX}} is a variational algorithm for real-time quantum simulation. It prescribes a succession of parameters $\vec\theta_t$ to produce ansatz states
}
\begin{align}
    \ket{\psi}_t = \hat{U}(\vec\theta_t) \ket{\text{in}},
\end{align}
\change{
which approximate the states produced from unitary time evolution of input state $\ket{\text{in}}$ under some Hamiltonian $\hat{H}$;
}
\begin{align}
    \ket{\psi}_t \approx \exp(- i \hat{H} t) \ket{\text{in}}.
\end{align}
\change{
In this manuscript, we merely use Li's method to generate interesting input states to our recompilation strategy. 
We include a more thorough description of Li's method in Appendix~{\ref{app:li_algorithm}}, including a short comparison with Trotterisation. See Reference~{\cite{Yuan2019}} for a detailed summary of variational quantum simulation techniques.
}

\subsection{Imaginary time evolution}

\change{
Imaginary time evolution is a variational algorithm for performing energy minimisation~{\cite{imag2018}}, or VQE~{\cite{suguru18}}. It requires the evaluation of a metric object that reconciles between parameter space and Hilbert space, recently shown to be an instance of quantum natural gradient{~\cite{natural_gradient}}.
It is incidentally a Wick rotation of the equations deriving Li's method, and drives the ansatz state toward the ground-state of the simulated Hamiltonian;
}
\begin{align}
    \ket{\psi}_t \approx N(t) \exp(- \hat{H} t) \ket{\text{in}},
\end{align}
\change{
where $N(t) = 1/\sqrt{\braket{\text{in}|\exp(-2 \hat{H} t)|\text{in}}}$ is a normalisation factor.
}
\change{
In this work, we use imaginary time evolution in our noise-free simulations as a subroutine to drive an ansatz state into the ground-state of a constructed Hamiltonian. In principle, any generic variational minimisation technique can be employed. However, imaginary time has been found in earlier studies to perform favourably for parameter optimisation}~\cite{imag2018,suguru18,Yuan2019}, \change{and exhibits good resilience to simple decoherence and shot noise}~\cite{imag2018}. 
\change{Additionally, it was recently shown to require only marginally more sampling than simple gradient descent}~\cite{balintNGcost}.
\change{
An elaboration of the role of imaginary time evolution is given in Appendix~{\ref{app:imag-time}}
}.

\subsection{Noisy natural gradient}
\label{noisyNatGrad}

\change{
For simulations of our recompilation algorithm on noise-burdened quantum computers, we replace imaginary time minimisation with an improved method for noisy settings. Recently, the metric evaluated in quantum natural gradient~{\cite{natural_gradient}} was adapted for non-unitary ansatzes, using the quantum Fisher information matrix~{\cite{noisy_natural}}.
Moreover, in that work it was established that for the case of noisy circuits producing output $\rho$, the elements $F_{k,l}$ of the metric can be well-approximated by 
}
\begin{align}
F_{k,l} = \text{Tr}[(\partial_k\rho)(\partial_l\rho)],
\end{align}
\change{
a quantity that can be evaluated experimentally via a swap-test circuit~{\cite{swaptest_for_overlap}}.
We will employ this generalisation of imaginary time evolution to the non-unitary case later, in the context of compilation with a noise-burdened quantum computer.
}

\section{Overview of compilation}

Consider a quantum circuit $\mathcal{A}$ acting on an $n$-qubit register and involving some arbitrary set of $m$ unitary gates $G^A_i$, each acting on one or more of the qubits. When the input state to the circuit is $\ket{\text{in}}$, the output is 
\begin{align}
\ket{\text{out}}=G_m^AG_{m-1}^A\dots G_1^A\, \ket{\text{in}}.
\end{align}

Now suppose that we wish to find an alternative circuit $\mathcal{B}$ which has the same, or nearly the same, action on the input state. This new circuit acts upon the same $n$-qubit register but generally may contain a different number $m^\prime$ of gate operations $G_i^B$. These new gates may be a very different set from those in $\mathcal{A}$, even (if we wish) a set that is too restrictive to express the former set exactly. 

The approach described here involves first compiling to a circuit $\mathcal{B}^{-1}$, which we define as the gate-by-gate inverse of $\mathcal{B}$. A comparable technique has been used for the related problem of learning an unknown state~\cite{Lee2018}, where one seeks a mapping to a known target in order that the reverse process can define the original state. We write the individual gates as $g^B_i=(G^B_i)^{-1}$ so that our circuits are,
\begin{align}
\mathcal{B}\rightarrow& G_{m^\prime}^B\dots G_2^BG_1^B\nonumber\\
\mathcal{B}^{-1}\rightarrow& g_1^Bg_2^B\dots g_{m^\prime}^B.\nonumber
\end{align}

Importantly, the gates $G^B_i$ are parameterised: each gate takes a single parameter $\phi_i$, such that $G^B_i(\phi_i=0)$ is the identity. Thus each $G^B_i$ is really a continuous family of unitaries. An example would be $\exp(i\phi\, \sigma_x^a\sigma_x^b)$ acting on qubits $a$ and $b$. It follows that each inverse gate $g^B_i$ is also a function of $\phi_i$. We are implicitly assuming that the gates are sufficiently simple (e.g. low enough qubit counts) that the mapping between $G^B_i$ and $g^B_i$ is tractable. Moreover, for the case that the recompilation process is using a quantum computer (as opposed to an all-classical implementation) we are implicitly assuming that $G^B_i(\phi_i)$ and  $g^B_i(\phi_i)$ are physically implementable for all values of $\phi_i$. (One could tolerate certain restrictions, but we do not explore that here).

To recompile $\mathcal{A}$, the circuit is applied to the input $\ket{\text{in}}$ and then circuit $\mathcal{B}^{-1}$ is applied, ultimately producing state $\ket{\text{fin}}$. 
\begin{equation}
\ket{\text{fin}}=\left(g_1^Bg_2^B\dots g_{m^\prime}^B\right)
\left(G_m^AG_{m-1}^A\dots G_1^A\right)\, \ket{\text{in}}.
\label{eqn:gG}
\end{equation}
The compiler seeks to find a set of parameter values $\vec{\phi}$ such that (as nearly as possible),
\begin{equation}
\ket{\text{in}}=\mathcal{B}^{-1}(\vec{\phi})\,\mathcal{A}\ket{\text{in}}
\ \ \ \Rightarrow\ \ \ \ 
\mathcal{B}(\vec{\phi})\,\ket{\text{in}}=\mathcal{A}\ket{\text{in}},
\label{eqn:theGoal}
\end{equation}
where it is understood that these equations are up to a meaningless global phase.

Initially $\phi_i=0$, $\forall i$ so that all gates $g^B_i$ are simply the identity. As the `user' we specify a {\it circuit template} for $\mathcal{B}$, since we fix the gate types and the sequence, but this template is `blank' i.e. free of parameter information. 

Generally finding satisfying $\vec{\phi}$ is a hard search problem since there may be thousands of parameters even for NISQ-era machines. One must therefore select the strategy carefully, giving consideration to potential problems such as becoming `stuck' in a local minimum as we evolve the parameter set. There are also issues relating to the device size: ideally recompilation will be achieved without the need for additional qubits.

The approach we take here coopts recent ideas relating to finding the ground state energy of a Hamiltonian. The device size remains $n$ qubits, and moreover although the scaling and performance of such approaches are not fully understood there is a developing literature on these topics~\cite{chemReview2018,McClean2016,Yuan2019}.

\subsection{Compilation by energy minimisation}

As noted in our list of restrictions, we are assuming that we understand  the specific input state $\ket{\text{in}}$ sufficiently to create a Hamiltonian for which that state would be the unique ground state. We stress that this Hamiltonian does not correspond to any real physical system of interest, it is  a fictitious construct purely to enable the recompilation process. We denote it $H_{\text{rec}}$ where the subscript stands for `recompilation'. Finding $H_{\text{rec}}$ is of course trivial for any input that has a product form: for example if $\ket{\text{in}}=\ket{00\dots 0}$ then the obvious choice would be $H_{\text{rec}}=\sum_j \sigma^z_j$. 

It may be desirable to ensure that there is a well-defined gap to the first excited state that is itself only $n-$fold degenerate (or to break such degeneracy if we wish). For the example just given, the gap is unity regardless of $n$. These properties are helpful in terms of the efficiency of the ground state finding protocol. 

Given that we have selected a suitable $H_{\text{rec}}$, then the recompilation process has become an eigensolving task: 

\vspace{4pt}
\parbox{.96 \columnwidth}{
\setlength{\parindent}{0cm}
\textit{
        Given a (fixed) input $\mathcal{A}\ket{\text{in}}$ to `ansatz' circuit $\mathcal{B}^{-1}(\vec{\phi})$, find the parameter values $\vec{\phi}$ for which the output has the lowest possible energy with respect to $H_{\text{rec}}$.
}
}
\vspace{4pt}

\change{
We can adopt any one of several quantum minimisation techniques to solve this problem~{\cite{chemReview2018}}. As motivated in Section~{\ref{sec:vari_sim_overview}}, we employ imaginary time evolution~{\cite{imag2018}} for our noise-free simulations, and a non-unitary adaptation~{\cite{noisy_natural}} for our noisy simulations.
We outline these methods in Appendices~{\ref{app:li_algorithm}} and {\ref{app:imag-time}}. 
}

Any variational eigensolving technique may become slow to evolve in specific cases. Anticipating this problem, we have explored a solution involving a series of proximal targets which in effect `lure' the process toward the eventual target. 
\change{
We present the technique in a following section, and offer an intuitive example in Appendix~{\ref{app:lure}}
}.
Note the specific demonstration of compilation in Section~\ref{sec:numerical_demo} does not in fact require such a lure (we do not observe any slowdown issues).

In the case that the compilation process is being performed with a quantum computer (rather than a classical emulation of the process), one may wonder whether the user would be able to determine how successful the compilation has been. A reasonable measure is the 
 fidelity between the input state $\ket{\text{in}}$ and its attempted reconstruction $\mathcal{B}^{-1} \mathcal{A} \ket{\text{in}}$, equivalent to that between $\mathcal{A}\ket{\text{in}}$ and $\mathcal{B}\ket{\text{in}}$. We report this measure from our simulations, however in the lab a user may not be able to evaluate this directly. Fortunately one can lower-bound the fidelity using the expected energy $\braket{H_{\text{rec}}}$ which is  measurable (indeed the imaginary-time variational approach involves repeatedly estimating quantities of this kind). Since $\ket{\text{in}}$ is the ground state of $H_{\text{rec}}$ with energy $E_0$, and presuming that we `understand' our fictitious Hamiltonian $H_{\text{{\text rec}}}$ sufficiently to know its first excited state energy $E_1$, then
\[
    \min\{\braket{H_{\text{rec}}}\}-E_0 = (1-F)(E_1-E_0)
\]
where the $\min$ denotes `minimum observable value given that the fidelity is $F$'. It follows that
\begin{equation}
    F \ge \frac{ E_1-\braket{H_{\text{rec}}}}{E_1 - E_0}.
    \label{eqn:fidAndEnergy}
\end{equation}

Note that if, as in the example above, our fictitious Hamiltonian $H_{\text{rec}}$ has a gap $E_1 - E_0$ of unity then the accuracy with which we can bound $F$ simply depends on the shot noise in our estimate $\braket{H_\text{rec}}$.

\subsection{The \textit{Lure} method}

\change{
The present task of mapping the state $\mathcal{B}^{-1}(\vec{\phi})\, \mathcal{A}\ket{\text{in}}$  `back' to state $\ket{\text{in}}$ can made more robust by seeking a series of intermediate mappings. In effect, we `lure' $\vec{\phi}$ towards the eventual goal incrementally. To achieve this, we generalise each gate $G_i^A$ in our original circuit $\mathcal{A}$ by introducing a global parameter $\alpha$ such that $G_i^A(\alpha=0)$ corresponds to the identity, and $\alpha=1$ produces the original gate in its full effect. 
}

\change{
As a single global variable common to all gates within $\mathcal{A}$, $\alpha$ then interpolates between the identity circuit $\mathcal{A}_{\alpha=0} = \mathbb{1}$, and the full circuit $\mathcal{A}_{\alpha=1}$.
We step $\alpha$ in small increments, at each step running the imaginary-time protocol to find a parameter set $\vec{\phi}$ that produces a sufficiently low energy state $\mathcal{B}^{-1}(\vec{\phi}) \mathcal{A}_{\alpha} \ket{\text{in}}$ according to our constructed Hamiltonian $H_\text{rec}$. We increment $\alpha$ by $\delta \alpha$ whenever we come within hyperparameter $\delta E$ of the ground state energy $E_0$. The final step sees $\alpha=1$ and the parameters $\vec{\phi}$ converge to their final values, which should `undo' the action of $\mathcal{A}$ on $\ket{\text{in}}$.
Other variant algorithms where $\alpha$ develops by a continuous change should have a similar performance. 
}

\change{
An example implementation of this lure method is provided in Appendix~{\ref{app:lure}}. 
Despite its simplicity, luring can markedly improve the convergence of the minimisation subroutine of recompilation.
The method was \textit{not} necessary for the simulations performed in Sec.~{\ref{sec:numerical_demo}}, but \textit{was} in an anomalous simulation performed in Sec.~{\ref{sec:numerical_scaling}}
}
\change{We note that the lure technique has some similarity to the incrementally evolving Hamiltonian employed in the adiabatic VQA demonstration of Ref.~{\cite{adiabaticVQAChen2020}}, however there the gradual adaption is to the cost function rather than the input state as here.}

\subsection{Adaptive timestep}
\label{sec:adaptive_timestep_intro}

\change{
In variational minimisation techniques like imaginary time evolution, the timestep ~$\Delta t$ is often a fixed hyper-parameter. Indeed in the proceeding Section~{\ref{sec:numerical_demo}}, $\Delta t$ will be chosen and fixed using a simple stability heuristic (as in~{\cite{imag2018}}).
However, an adaptive timestep algorithm can scale this hyper-parameter during minimisation, to significantly speed-up convergence in drastically changing landscapes. This becomes important in increasingly large Hilbert spaces, where motivating a fixed parameter becomes difficult, like the during the scaling tests in Section~{\ref{sec:numerical_scaling}}.
}
\change{
We here present a custom adaptive timestep algorithm based on line search~{\cite{linearsearch1986}} which sees the timestep automatically change through orders of magnitude during optimisation.
}

\change{
Assume $\Delta t$ was the timestep used in a particular iteration of parameter optimisation (e.g. imaginary time evolution). The next iteration involves computing the energy gradient $\nabla \braket{E(\vec\theta)}$, solving a linear equation for $\frac{\mathrm{d}\vec\theta}{\mathrm{d}t}$, and updating the parameters according to
$\vec\theta \to \vec\theta +   \Delta t \frac{\mathrm{d}\vec\theta}{\mathrm{d}t} $.
}
\change{ 
In general, experimentally measuring the energy gradient $\nabla \braket{E(\vec\theta)}$ is a task that scales linearly with the number of parameters, unlike measuring the energy at a point, $\braket{E(\vec\theta)}$.
Hence, after computing $\nabla \braket{E(\vec\theta)}$, we can afford to make several additional energy measurements at little relative total cost. It is worthwhile to use the current gradient to produce several choices of $\Delta t$, measure their associated energies, and pick the best. This is advantageous because the energy landscape in parameter space can itself vary drastically, so a fixed $\Delta t$ can become unsuitable.
}
\change{
This is the basis of our algorithm, which we present formally in Algorithm~{\ref{alg:adapting_timestep}}
}.

\begin{algorithm}
\caption{Adaptively updating $\Delta t$ each iteration of parameter minimisation.}
\label{alg:adapting_timestep}

\textbf{Require:} $\Delta t$ is the previous iteration's timestep

\textbf{Require:} $E_0$ is the previous iteration's energy

\textbf{Require:} $\vec{\theta}$ are the current parameters and $\frac{\mathrm{d}\vec\theta}{\mathrm{d}t}$ their pre-computed derivatives

\texttt{\\} 

\textcolor{orange}{// energy of nominated timestep under current gradient}

\textbf{define} \textcolor{blue}{EN}($\Delta t$)

\Indp

    \textbf{return} $\braket{E(\vec\theta + \Delta t \frac{\mathrm{d}\vec\theta)}{\mathrm{d}t}}$

\Indm

\texttt{\\} 

\textcolor{orange}{// returns adapted timestep}

\textbf{define} \textcolor{blue}{TIMESTEP}($\Delta t$, $\epsilon_{\Delta t/2}$,
    $\epsilon_{\Delta t}$,
    $\epsilon_{2 \Delta t}$
)

\Indp

    $\epsilon_{\text{min}} = $ \text{min}($\epsilon_{\Delta t/2}$,
    $\epsilon_{\Delta t}$,
    $\epsilon_{2 \Delta t}$)
    
    \texttt{\\} 
    
    \textcolor{orange}{// halt when converged}
    
    \textbf{if} $|\epsilon_{\text{min}}| \le 10^{-8}$
    
    \Indp
    
        \textbf{return} $\Delta t$
        
    \Indm
    
    \texttt{\\} 
    
    \textcolor{orange}{// shrink rapidly when $\Delta t$ is unstably large}
    
    \textbf{if} $\epsilon_{\text{min}} > E_0$
    
    \Indp
        
        \textbf{return} \textcolor{blue}{TIMESTEP}($\Delta t/8$, \textcolor{blue}{EN}($\Delta t/16$), \textcolor{blue}{EN}($\Delta t/8$), \textcolor{blue}{EN}($\Delta t/4$))
        
    \Indm
    
    \texttt{\\} 
    
    \textcolor{orange}{// adjust $\Delta t$ until optimal, re-using previous energies}
    
    \textbf{if} $\epsilon_{\text{min}} = \epsilon_{\Delta t}$
    
    \Indp
    
        \textbf{return} $\Delta t$
    
    \Indm
    
    \textbf{if} $\epsilon_{\text{min}} = \epsilon_{\Delta t/2}$
    
    \Indp
    
        \textbf{return} \textcolor{blue}{TIMESTEP}($\Delta t/2$, \textcolor{blue}{EN}($\Delta t/4$), $\epsilon_{\Delta t/2}$,
        $\epsilon_{\Delta t}$)
    
    \Indm
    
    \textbf{if} $\epsilon_{\text{min}} = \epsilon_{2 \Delta t}$
    
    \Indp
    
        \textbf{return} \textcolor{blue}{TIMESTEP}($2 \Delta t$, 
        $\epsilon_{\Delta t}$,
        $\epsilon_{2 \Delta t}$
        \textcolor{blue}{EN}($4 \Delta t$))
    
    \Indm
    
\Indm 

\texttt{\\} 

\textcolor{orange}{// adapt timestep, starting from previous}

$\Delta t = $ \textcolor{blue}{TIMESTEP}($\Delta t$, 
        \textcolor{blue}{EN}($\Delta t/2$)
        \textcolor{blue}{EN}($\Delta t$)
        \textcolor{blue}{EN}($2\Delta t$)
        )

\end{algorithm}
%
%
%

\change{
Via Algorithm~{\ref{alg:adapting_timestep}}, $\Delta t$ can be enlarged exponentially quickly if the current $\vec\theta$ lies on a persistent slope, and shrunk exponentially quickly if $\Delta t$ has become unstably large.
It includes the circumstance whereby no choice of $\Delta t/2$, $\Delta t$ or $2 \Delta t$ decrease the energy, implying that $\Delta t$ is unstably large, since it is known that a sufficiently small timestep of variational imaginary time evolution should exhibit a monotonic decrease in energy~{\cite{imag2018}}. 
}

\subsection{Further gate elimination}
\label{sec:gate_elim_intro}

After recompiling $\mathcal{A}\ket{\text{in}}$ into $\mathcal{B}(\vec{\phi}) \ket{\text{in}}$ using the above methods, we can attempt to further shrink the circuit by eliminating gates with small parameters (thus deviating from the user-specified template).
We choose the parameter $\phi_j$ whose current value $\phi_j=\delta$ is closest to $0$ (or more strictly, since we may be dealing with periodic functions, we identify $j$ for which $g^B_j(\phi_j)$ is closest to the identity). We then continue our imaginary-time evolution under modified variational equations, where we additionally constrain that within each iteration,
\begin{align}
    \Delta \phi_j = - \frac{\delta}{ N }.
\end{align}
Here $N$ limits the change in $\phi_j$ in a single iteration. This simultaneously drives $\phi_j$ toward zero while retaining the pressure toward the ground state. Generally we will reach ${\phi}_j=0$ having suffered a small penalty in energy (and thus fidelity of the new circuit). Once zero is reached, we remove gate $g^B_j = \mathbb{1}$ and then repeat the process. This continues until the energy has unacceptably risen. What is `unacceptable' will depend on the application, but for the examples here we set the threshold to be twice the gap between true ground and the original energy of $\mathcal{B}^{-1} \mathcal{A} \ket{\text{in}}$ under $\hat{H}_{\text{rec}}$. \change{In other words, we permit the energy defect to double in return for reducing the number of gates in the output circuit. We found this threshold enabled excellent circuit compression despite only a modest reduction in fidelity}. We emphasise that this entire phase is an optional post-process after the main recompilation. We will denote the resulting circuit of this additional gate elimination process as $\mathcal{B}_{\text{elim}}$.

\section{Numerical demonstration}
\label{sec:numerical_demo}

Having thus described the compilation and optimisation process in general terms, we now illustrate it with a specific example: recompilation of a $7$-qubit, $186$-gate circuit into a quite different template.

\subsection{Selecting an interesting example circuit \texorpdfstring{$\mathcal{A}$}{A}}
\label{sec:SpecifyingA}

\begin{figure}[!htbp]
\centering{}
\includegraphics[width=0.9\columnwidth]{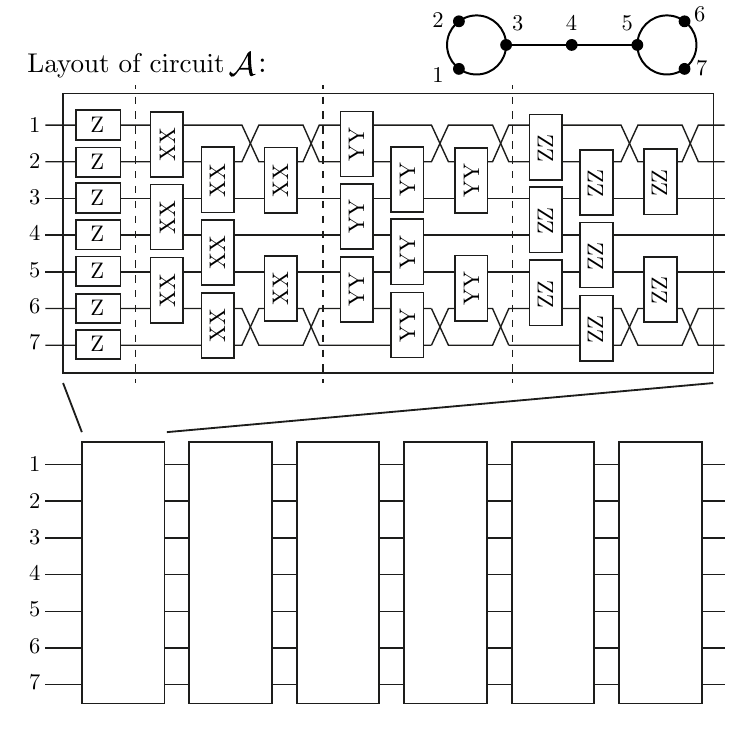}
\caption{
The circuit $\mathcal{A}$ which we opt to use as the input to our compilation process. The upper right figure summarises the two-qubit gate connections. The circuit is related to a quantum dynamics problem as described in Section~\ref{sec:SpecifyingA}, but for recompilation purposes one can regard it as an arbitrary pattern of 186 unique non-Clifford gates (including 144 two-qubit gates). Here $Z(\theta)=\exp(-i\frac{\theta}{2} \sigma^Z)$, $ZZ(\theta)=\exp(-i\frac{\theta}{2} \sigma^Z\otimes \sigma^Z)$, and similarly for the $Y$ and $Z$ gates. The angle $\theta$ is unique for each gate, and for completeness we specify the values in a table in the appendix. 
}
\label{fig:circuitA}
\end{figure}

This section describes how we select an interesting and complex $\mathcal{A}$ circuit as the object that we will attempt to recompile into a new form. In essence: we choose a $7$-qubit circuit relevant to a certain simulation task and specify it in Fig.~\ref{fig:circuitA}; we also choose initial state $\ket{\text{in}}=\instate$. {\it Readers who are concerned only with the recompilation process may care to skip the rest of this section.}

Rather than randomly generating the circuit $\mathcal{A}$, we focus on the likely application areas for our recompilation technique: hybrid algorithms that aim at dynamical simulation or eigensolving. Given that the recompilation technique itself involves a kind of eigensolver, for clarity we opt instead to make the circuit $\mathcal{A}$ relevant to a dynamical simulation task. Specifically, we assume that we wish to model the evolution of a certain $7-$spin network, with the topology of spin-spin interactions shown in the upper right of Fig.~\ref{fig:circuitA}. We take it that the Hamiltonian of this system is 
\begin{equation}
H_{\text{sys}}=\sum_i B_i \sigma_i^z + 
\sum_{i,j} \sum_{S\in x,y,z}J_{i,j}^S \sigma^S_i\sigma^S_j
\label{eqn:spinH}
\end{equation}
where $\sigma$ are the standard Pauli operators and the constants $B_i<0$ and $J^S_{i,j}>0$ as listed in Appendix~\ref{app:numerical_sim_details}. This is therefore a rather general spin network with irregular antiferromagnetic interactions and local fields. In order to create an interesting evolution we select the initial state $\Psi(0)=\instate$ i.e. a product state where one qubit is orientated such that it has maximum energy with respect to its local field and the others have zero expected energy in their local fields. 
We choose a simple recompilation Hamiltonian for which $\Psi(0)$ is the ground state, namely
\[
\hat{H}_{\text{rec}} = 
\sigma^z_1 - \sum\limits_{j=2}^7 \sigma^x_j.
\]

As a relevant test of our recompilation technique, we stipulate that the purpose of original circuit $\mathcal{A}$ is to model the evolution of this system, i.e. to create (a good approximation to) the state
\[
\ket{\Psi(t)}=\exp(i H_{\text{sys}}t)\,\ket{\Psi(0)}
\]
for some time $t$ which we presently specify.
A naive approach might be to use a number of \change{Trotter `cycles'}~\cite{trotterformula,trotterforsim} i.e. to use a number $q$ of identical circuit blocks each of which contains one gate for each Pauli term in $H_{\text{sys}}$. Each gate $\mathcal{T}_j$ within the first Trotter cycle would therefore be of the form 
\[
\mathcal{T}_j=\exp\left(-i\frac{\theta_j}{2}K_j\right)
=\left(\cos\frac{\theta_j}{2}\right)\,\mathcal{I}-i\,\left(\sin\frac{\theta_j}{2}\right)\,K_j
\]
where $j$ runs $1\dots 31$ in our case. Here $K_j$ is the Pauli operator from the $j^{\text{th}}$ term of $H_{\text{sys}}$, i.e. either a single or double $\sigma$ operator. Meanwhile $\theta_j/2=C_jt/q$ where $C_j$ is the constant in the $j^{\text{th}}$ term of $H_{\text{sys}}$, i.e. a $B$ or $J$ value. Thus for $\mathcal{T}_1$ we would use $K_1=\sigma_z$ and $\theta_1=2B_1\,t/q$. Gates deeper into the circuit each replicate a gate in the first cycle, i.e. $\mathcal{T}_{j+31}=\mathcal{T}_j$. For sufficiently small $t$ this approach is guaranteed to provide an accurate simulation.
Fig.~\ref{fig:circuitA} shows a circuit of this kind. 

Following this basic Trotter rule for selecting the $\theta_j$ values leads to rather too simple a structure to test our recompilation protocol fully because of the gate recurrence $\mathcal{T}_{j+31}=\mathcal{T}_j$. (As an aside we remark that we have verified this: Setting the circuit's $\theta$ values to correspond to $t=0.75$ where its simulation fidelity is $0.9983$, we find we can recompile to a circuit with only about half the number of two-qubit gates and yet still retain simulation fidelity above $0.998$). The ease of recompiling this standard Trotter circuit is related to the fact that it does not make optimal use of the gates available. Given the same gate layout, as shown in Fig.~\ref{fig:circuitA}, one can instead use the variational algorithm described in Ref.~\cite{YingPRX} to adjust the `strength' $\theta_j$ of each gate independently of all others in an optimal fashion. We indeed apply this algorithm, which we introduced earlier as Li's algorithm and we describe more completely in Appendix~\ref{app:li_algorithm}, to create our circuit $\mathcal{A}$. We choose the time $t=1.75$ as this is toward the outer limit of the range for which the circuit structure in Fig.~\ref{fig:circuitA} can produce an accurate simulation using Li's algorithm. The resulting parameters $\theta_j$ are specified in Table~\ref{tab:state_before_recomp} in Appendix~\ref{app:numerical_sim_details}; configured this way, the circuit in Fig.~\ref{fig:circuitA} successfully replicates the state of the simulated spin system at time $t=1.75$ with a fidelity of $0.995$.

Although the purpose here was simply to create a complex but meaningful circuit for recompilation, in doing so we did make a number of interesting observations about the power of Li's algorithm as compared to Trotter approaches. For the interested reader these are described in Appendix~\ref{app:li_algorithm}.

\subsection{Performance of the recompilation}
\label{performance_subsection}
\begin{figure}[!htbp]
\centering{}
\includegraphics[width=1\columnwidth]{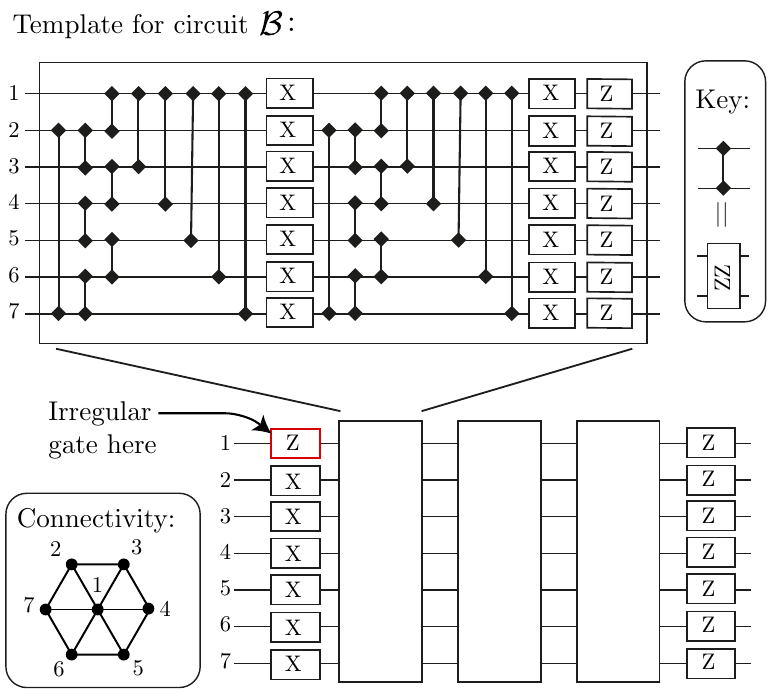}
\caption{
The template for the recompiled circuit $\mathcal{B}$. The template is user-specified, and the recompilation process will determine the $\phi$ value for each gate. The template's structure is quite different to the original circuit: The template has a smaller family of gate types ($YY$ and $XX$ type gates are omitted), it has half as many two-qubit gates in total (72 rather than 144), but a larger number of single-qubit gates (77 versus 42). Moreover the topology of the two-qubit gates is different: it is a triangular lattice forming a hexagon as shown in the inset.
}
\label{fig:circuitB}
\end{figure}

\begin{figure}[!htbp]
    \centering
    \includegraphics[width=\columnwidth]{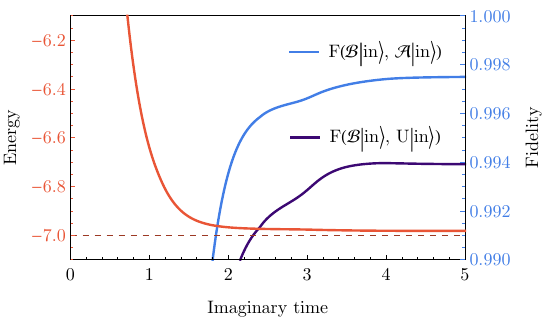}
    \caption{The process of recompiling circuit $\mathcal{A}$ into the template $\mathcal{B}$. We use an imaginary time variational algorithm~\cite{imag2018} as described in Appendices~\ref{app:li_algorithm} and \ref{app:imag-time}, according to which the $\phi_i$ parameters evolve through an iterative process until no further reduction in $\braket{H_{\text{rec}}}$ occurs. 
    %
    The light blue curve shows the fidelity with which $\mathcal{B}\ket{\text{in}}$ matches the target $\mathcal{A}\ket{\text{in}}$. Since the fidelity is not directly experimentally measurable, we also plot the user measurable quantity $\langle H_{\text{rec}}\rangle$ in red, which bounds the fidelity as stated in Eqn.~(\ref{eqn:fidAndEnergy}). (Also shown in dark purple is the fidelity with respect to the true state of the simulated spin system, see Section~\ref{sec:SpecifyingA}. The circuit $\mathcal{A}$ itself has a finite simulation fidelity of $0.995$ and thus the recompiled circuit is somewhat lower.)}
    \label{fig:ansatz_recomp_first}
\end{figure}

To make the recompilation task interesting, we specify a template for $\mathcal{B}$ which differs from $\mathcal{A}$ in several ways: Firstly, we select a different set of gates. The original circuit $\mathcal{A}$ involves 6 types of gate (single qubit rotations about the $X$, $Y$ and $Z$ axes, and two qubit rotation gates around $XX$, $YY$ or $ZZ$). For the new template $\mathcal{B}$ we opt to use the same set of single-qubit gates but we restrict ourselves to only the $ZZ$-based two-qubit gate. (A restriction of this kind is relevant to real devices which typically have a native type of two-qubit gate that is the least onerous to perform, and therefore it is natural to attempt to recompile into a template featuring only one kind of two-qubit gate). 
We also vary the connectivity: instead of mimicking the Trotter terms of the Hamiltonian $H_{\text{sys}}$, as in circuit $\mathcal{A}$, we now adopt a centred hexagon as shown in Fig.~\ref{fig:circuitB}. 

The recompilation process then proceeds as described earlier. The task of the classical computer, i.e. the solution of the simultaneous equations which yields the appropriate parameter updates at each step, is performed via truncated singular value decomposition as described in Appendix~\ref{app:numerical_sim_details}. Figure~\ref{fig:ansatz_recomp_first} shows how the parameters develop. The upper part of this figure indicates the performance of the recompilation process; reaching the target energy of $\braket{H_{\text{rec}}}=-7$ would indicate perfect recompilation. Also shown is the fidelity of $\mathcal{B}\ket{\text{in}}$ versus $\mathcal{A}\ket{\text{in}}$, which the user would not have access to. The final fidelity of the process is $0.998$. Given the restricted nature of the template, and particularly the fact that it has only half as many two-qubit gates, the recompile is remarkably effective.

We remark that, in other examples that we have studied we have used a template with a higher gate count and recompiled to (essentially) perfect fidelity: infidelity of order $10^{-5}$ has been observed, and it seems probable that this is non-zero only because of imperfections in the numerical solution methods. We anticipate that these methods can be further refined.

\begin{figure}
    \centering
    \includegraphics[width=\columnwidth]{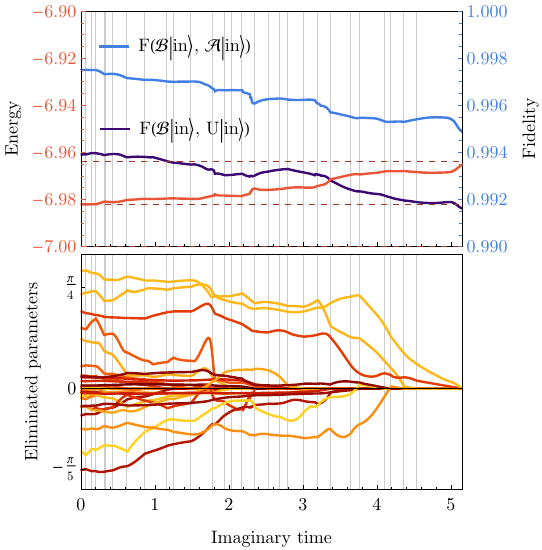}
    \caption{
    The gate elimination post-process, following the recompilation stage. We require certain of the $\phi_j$ parameters to diminish to zero during additional iterations of imaginary time evolution, so that the corresponding gate can be removed. This continues until the energy $\braket{H_{\text{rec}}}$, shown in red in the top panel, deviates from the known minimum (here, $-7$) by some chosen threshold (here, double the oroginal defect). The lower panel shows how the corresponding parameters are eliminated; in total thirty are set to zero, reducing the circuit size from $149$ to $119$. Vertical lines indicate each time a parameter is eliminated.
    \label{fig:gate_elim}
    }
\end{figure}

As a post-recompile stage, we apply the gate elimination process described in Section~\ref{sec:gate_elim_intro}. It is necessary to select a tolerance for the process, i.e. a level of reduction in the overall quality of the recompiled circuit which we are prepared to tolerate in return for further `compressing' the circuit. For the present example, we assume that we will tolerate a doubling in the energy defect with respect to the ideal value of $-7$. The performance of this process is indicated in Fig.~\ref{fig:gate_elim}. Remarkably we find that we can eliminate a further 30 gates (11 single-qubit, 19 two-qubit) in this fashion, and that the fidelity of the resulting circuit is still high at $0.995$. The final circuit includes only $119$ gates (whereas the original $\mathcal{A}$ has $186$) and moreover the number of two-qubit gates has been reduced almost to a third (from 144 to 53). Recall that all remaining two-qubit gates are all of a single type, $ZZ=\exp(-i \frac{\theta}{2} \sigma^Z \otimes \sigma^Z)$.

We emphasise that these recompilation fidelities have been achieved with a non-trivial circuit $\mathcal{A}$ that is {\it already} optimised, with respect to a simple Trotter circuit, through the use of Li's algorithm. As noted earlier, applying our recompilation to a simple Trotter circuit will produce more dramatic results, e.g. compression to $\sim50\%$ depth with only a very small loss of fidelity ($\Delta F=0.0002$ in that case). 

\begin{figure}
    \centering
    \includegraphics[width=\columnwidth]{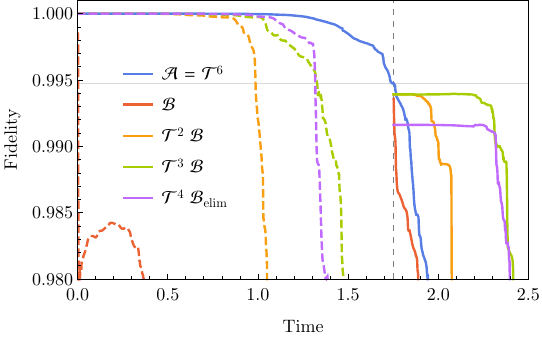}
    \caption{
    Realtime simulation of the spin system specified in Eqn.~(\ref{eqn:spinH}) from $t=0$ to $t=2.5$ using Li's algorithm~\cite{YingPRX} with various possible circuits. In blue, the original structure $\mathcal{A}$, which we previously `froze' at $t=1.75$. We note if we allow its parameters $\theta$ to continue to evolve using Li's algorithm, past $t=1.75$ the fidelity of the simulation drops precipitously. In red, the recompiled circuit $\mathcal{B}$ is seen to fare somewhat worse than when its parameters $\phi$ are similarly evolved by Li's algorithm. 
    However, the green line corresponds to the performance of the recompiled and augmented circuit (as described in the main text) and is superior: it can sustain high fidelity simulation for a further $~0.4$ time units. We highlight the resource cost of the featured circuits in Fig.~\ref{fig:resource_count}.}
    \label{fig:returningToSim}
\end{figure}

This concludes our description of our recompilation example. However, given the success of the process, we were interested to consider the following questions which relate to the nature of circuit $\mathcal{A}$ as a simulation task. As explained earlier in Section~\ref{sec:SpecifyingA}, the meaning of $\mathcal{A}$ is that it reproduces, with fidelity $0.995$, the state of a certain physical spin system (Eqn.~(\ref{eqn:spinH})) at time $t=1.75$. 

\noindent (1) If we take the template form of $\mathcal{B}$ with all $\phi_j=0$ and use it with Li's algorithm `in the first place' without ever considering the Trotter-inspired layout of $\mathcal{A}$, will we obtain a high performance simulation out to $t=1.75$? If this were the case it would rather obviate the need for recompilation, at least for this type of application. However the answer is `no'. As shown in Fig.~\ref{fig:Trotter_comparison} by the red-dashed line, the template $\mathcal{B}$ performs very poorly as a basis for the simulation. Its fidelity drops to $\sim0.98$ almost immediately. This is because a small increment in the parameters in  $\mathcal{B}$ does not correspond to a small shift in time for the simulated system (circuit $\mathcal{B}$ has the `wrong' topology, as shown by the inset in Fig.~\ref{fig:circuitB} versus the inset in Fig.~\ref{fig:circuitA}). To create our compact representation of the circuit that simulates the state of the spin system at $t=1.75$, we did indeed need to use a circuit whose structure reflects the Hamiltonian (Eqn.~(\ref{eqn:spinH})) to reach $t=1.75$ and only then recompile it.

\begin{figure}[tbph!]
    \centering
    \includegraphics[width=.675\columnwidth]{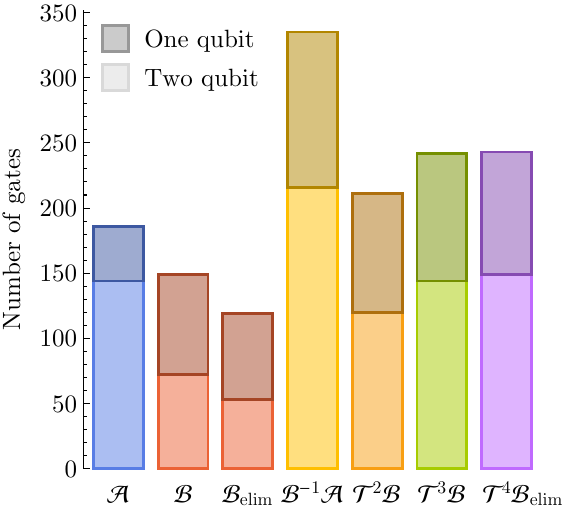}
    \caption{
    Resource count of the considered circuits in Fig.~\ref{fig:returningToSim}, and those involved in recompilation. Recall that $\mathcal{A}$ is the input circuit of 6 Trotter cycles though with parameters freely evolving under Li's method. $\mathcal{A}$ is recompiled into $\mathcal{B}$ via the use of ansatz $\mathcal{B}^{-1} \mathcal{A}$ in an imaginary-time extension of Li's method. At a further fidelity penalty, one can eliminate the weakest gates in $\mathcal{B}$ to obtain $\mathcal{B}_{\text{elim}}$. After recompiling, realtime simulation can be continued by appending additional Trotter cycles $\mathcal{T}$ to $\mathcal{B}$ or $\mathcal{B}_{\text{elim}}$ (as indicated in Fig.~\ref{fig:returningToSim}).
    }
    \label{fig:resource_count}
\end{figure}

\noindent (2) What happens if we now augment the (recompiled, compressed) circuit by `pasting on' a number of `blank' Trotter cycles so as to recover roughly the two-qubit gate count of the original circuit $\mathcal{A}$, and then proceed with the simulation which motivated that circuit? Substituting the new recompiled and augmented circuit, will the simulation of the spin system's dynamics using Li's algorithm proceed forward from $t=1.75$ with good fidelity? The results are shown in in Fig.~\ref{fig:returningToSim}. We see that this does indeed produce a superior performance versus simply continuing past the $t=1.75$ point with a circuit of the form of $\mathcal{A}$. \change{
Fig.~{\ref{fig:resource_count}} visualises the resource costs of the tested simulation strategies, colour coded to match their performances in Fig.~{\ref{fig:returningToSim}}. It highlights that the significant improvement of circuit $\mathcal{T}^3 \mathcal{B}$ over $\mathcal{A}$ in the future phase of real-time simulation was achieved without increasing the number of two qubit gates.
}

\section{Robustness to noise}
\label{sec:robustness}

\change{
The methods described in this paper are relevant to two scenarios: compiling a quantum circuit using a classical machine, as in the simulations here, or doing so with the use of an actual quantum computer to perform the optimisation. The latter is necessary for large circuits given that classical machines cannot efficiently emulate general circuits of more than about $50$ qubits. In scenarios where a quantum computer is available for the recompilation, it might be a fault-tolerant machine capable of practically perfect execution or it might be a NISQ-era machine with inherent noise in gates, measurements, etc. Indeed the latter is a likely scenario given that NISQ-era computers will have a critical need of optimised circuits. However the possibility of compilation with a noise-burdened system raises the important question of whether the output of the compilation, i.e. the classical parameters specifying the gates in the recompiled circuit, would be severely degraded by the noise in the compilation process. This is a question we address through a series of further simulations. 
}

\change{
An encouraging study~{\cite{resilience_of_variational}} by Sharma \textit{et al} notes that VQE-type optimisation of a cost function can be remarkably robust against noise in the system. Furthermore, the methods mentioned earlier in subsection~{\ref{noisyNatGrad}} provide a mechanism to formally incorporate noise (or indeed other non-unitary processes) into the imaginary descent mechanism which we have adopted throughout this paper. Specifically, in Ref.~{\cite{noisy_natural}} the metric required for imaginary time (or `quantum natural gradient') methods is shown to relate to quantum Fisher Information in the general case, and an experimentally-practical method to estimate that quantity is described. It is therefore possible, if computationally demanding, to investigate the generalisation of the cases considered earlier in this paper to scenarios involving circuit noise.
}

\begin{figure}
    \centering
    \includegraphics[width=\columnwidth]{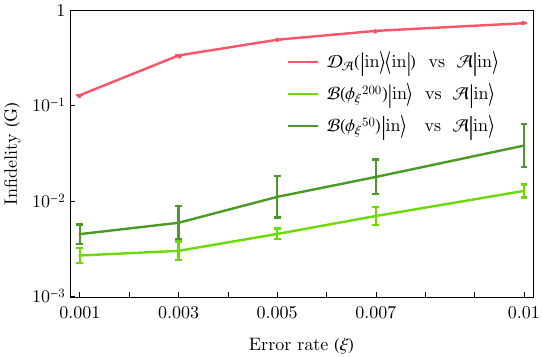}
    \caption{
    \change{
    Recompilation using a quantum computer that suffers significant noise. The task is that explored in the previous section, namely recompilation of $\mathcal{A}\ket{\text{in}}$ (Fig.~{\ref{fig:circuitA}}) into $\mathcal{B}(\vec{\phi})\ket{\text{in}}$ (Fig.~{\ref{fig:circuitB}}), though using the cost function given in Eq.~{\ref{eq:noisy_rec_hamel}}.
    The horizontal axis $\xi$ is the severity of the two-qubit depolarising noise following each two-qubit gate (single qubit gates receive severity $\xi/10$).
    The green lines show the infidelity between the target  state $\mathcal{A}\ket{\text{in}}$ and the pure state generated by $\mathcal{B}$ with parameters ${\phi_\xi}^n$, as learned from $n$ iterations of noisy quantum natural gradient. Specifically, this is $1 - \bra{\text{in}}\mathcal{A}^\dagger \mathcal{B}({\phi_{\xi}}^n)\ket{\text{in}}$, and is the error introduced by the noisy compilation process itself.
    The red line (uppermost) is included for context, and shows the infidelity between the ideal target state $\mathcal{A}\ket{\text{in}}$ and its noisy counterpart $\mathcal{D}_\mathcal{A}( |\text{in}\rangle\langle\text{in}|)$, where $\mathcal{D}_\mathcal{A}$ is the channel prescribed by inserting noise into $\mathcal{A}$. Concretely, the red line shows $1 - \bra{\text{in}}\mathcal{A}^\dagger \mathcal{D}_\mathcal{A}(|\text{in}\rangle\langle\text{in}|) \mathcal{A}\ket{\text{in}}$. One observes that the blue line is orders of magnitude lower than the red, confirming that recompilation is remarkably noise-robust. Generally $9$ independent runs were collated for each plotted value of $\xi$, except $\xi=0.01$ for which there were $14$ runs. 
   }
   }
   \label{fig:noisy_compile}
\end{figure}

\change{
We revisit the previous problem presented of recompilation from $\mathcal{A} \ket{\text{in}}$ (shown in Fig.~{\ref{fig:circuitA}}) into the format of circuit $\mathcal{B} \ket{\text{in}}$ (shown in Fig.~{\ref{fig:circuitB}}). As before, we seek the parameter set $\vec{\phi}$ that allows $\mathcal{B}$ to most closely match the action of $\mathcal{A}$ on a given input state (see Eqn.~{\ref{eqn:theGoal}}). The performance given error-free gates was described in subsection~{\ref{performance_subsection}} and Fig.~{\ref{fig:ansatz_recomp_first}}. We now introduce errors afflicting \textit{all} quantum gates, including those in both circuits (since both are required in the recompilation task). Our goal is still to find the optimal parameter set for the noise-free case, therefore we can expect that the noise in both circuits is deleterious for this task. 
}

\change{
We therefore modify our previous simulation by introducing per-gate noise, relevant error mitigation, and a modified cost function, as described below:
}
\begin{enumerate}
    
   \item
   \change{
   We apply depolarising noise subsequent to all initialisation and gate operations. We apply either single-qubit or two-qubit correlated noise depending on the gate that has just occurred. Two-qubit depolarising is of severity $\xi$ and acts on qubits $i,j$ with the mapping
   }
   \begin{align}
   \rho\rightarrow (1-\xi)\rho+\frac{\xi}{15}\sum_{k}N^k\rho N^{k\,\dagger},
   \end{align}
   \change{
   where $N^i$ are the fifteen operators $\mathcal{I}_i\sigma^X_j,\mathcal{I}_i\sigma^Y_j$,... though to $\sigma^Z_i\sigma^Z_j$. Meanwhile the single-qubit depolarising channel has the analogous form with just three operators $N$ and lower severity $\xi/10$. This reflects the fact that in experiments single qubit gates typically have higher fidelity.
   }
   
   \item  
   \change{
   In order to generalise the imaginary time (a.k.a. natural gradient) method to these non-unitary, noise-burdened gates, the metric tensor key to that approach must be generalised~{\cite{noisy_natural}}.
   In the lab, the evaluation of the metric requires a swap-test involving two copies of the circuit output together with an ancilla, implying $2\times 7+1=15$ qubits for the present case. Note that in our present density matrix simulations, both circuit outputs are represented identically going into the swap-test step, and can therefore employ a single copy and apply a generous degree of noise in modelling the swap-test process itself. Further details are in Appendix~{\ref{app:numerical_sim_details}}.
   }
   
   \item 
   \change{
   We presume that an experimental effort would  employ error mitigation to negate noise as far as possible, and so we incorporate such a method in our model. We adopt the extrapolation technique which is a well-established
   }~\cite{YingPRX,TemmePRL}
   \change{
   and experimentally validated~{\cite{IBMerrorMitiExp}} method with a simple protocol: Each observable required for the optimisation process is estimated by measuring at noise level $\xi$, and again at noise level $2\xi$ (in our simulations, this is done without the addition of further noise due to finite sampling, i.e. shot noise, so that values will correspond to the high-sampling limit). The recorded value is then the observable extrapolated to $\xi\rightarrow 0$ using the presumption of exponential decay~{\cite{suguruPRX}}.
   }
   
    \item 
    \change{
    We select a fictitious Hamiltonian with a cost function which we find efficient for this more computationally-demanding scenario:
    }
    \begin{align}
    H_\text{rec}=\mathcal{I}-\vert {\Psi(0)}\rangle\langle {\Psi(0)}\vert.
    \label{eq:noisy_rec_hamel}
    \end{align}
    \change{
    As before, $\ket{\Psi(0)}=\instate$. This cost function is straightforward to evaluate experimentally by measuring each qubit in the relevant basis. For problems beyond the 7-qubit case considered here, we would likely adopt a localised version of this global function, in line with recent work arguing for superior scaling via local functions~{\cite{cerezoNatComms2021}}.
    }
   
\end{enumerate}

\change{
We performed simulations for various values of the noise severity $\xi$ ranging between $0.1\%$ and $1\%$, presented in Fig.~{\ref{fig:noisy_compile}}. We emphasise $\xi$ is the noise for each two-qubit gate (with additional, weaker, noise on single-qubit gates). One might expect the compilation process to fail completely for $\xi$ approaching $1\%$ given that the total recompilation circuit has about $250$ gates (as it involves both circuits $\mathcal{A}$ and $\mathcal{B}$, shown in Figs.~{\ref{fig:circuitA}} and {\ref{fig:circuitB}} respectively). We note that even the original circuit $\mathcal{A}$ acting alone will produce a very noisy output with this level of gate imperfection; in Fig.~{\ref{fig:noisy_compile}} the solid red line shows the infidelity in the output of $\mathcal{A}$ versus the noise-free case, and we observe severe degradation (order $10\%$ infidelity) even for noise level $\xi=0.1\%$, rising above $70\%$ infidelity at the  $\xi=1\%$ level. 
}

\change{
Despite these high levels of noise, recompilation in fact succeeds remarkably well. Recall that the purpose of the recompilation is to determine the best set of classical parameters $\vec{\phi}$ with which to configure variational circuit $\mathcal{B}$ so when it acts on input $\instate$, its output most closely matches that of circuit $\mathcal{A}$ given the same input. Therefore to measure the effectiveness of the recompile we plot the infidelity
}
\begin{align}
G=1-\text{F}\left(\mathcal{B}(\vec{\phi_\xi})\ket{\text{in}},\mathcal{A}\ket{\text{in}}\right),
\end{align}
\change{
where circuits are now noiseless but circuit $\mathcal{B}$ is configured with the classical parameters $\vec{\phi_\xi}$ obtained from our (noisy, with severity $\xi$) recompilation process.
}

\change{
The solid blue line in Fig.~{\ref{fig:noisy_compile}} shows this infidelity $G$. We observe that even for severe noise at the $1\%$ per gate level, the recompile still produces good classical parameters $\vec{\phi_\xi}$. The use of $\vec{\phi}_{\xi=0.01}$ instead of the ideal $\vec{\phi}_{\xi=0}$ causes $G$ to rise from $0.3\%$ to about $1.1\%$, whereas the noise is now so severe that the output of any variant, even the original circuit $\mathcal{A}$, is now highly mixed. One can reasonably state that \textit{in circumstances where the noise is not too high for circuits to be of some value}, the recompilation remains useful and indeed remarkably accurate. In this sense, we conclude recompilation with noisy circuits is entirely practical, and the use of the advanced imaginary time optimisation method does not impede this. 
}

\change{
We should note two caveats to this encouraging conclusion: Firstly, that only the depolarising model has been studied so far, and secondly that there are increasing sample costs for increasing noise severity. For the former, while this is something that could be investigated comprehensively in a followup study, we feel confident that the very high degree of robustness suggests that other models will be at least moderately well-tolerated. We note that VQA approaches are inherently blind to systematic, deterministic errors (albeit one would need to correct the obtained-parameters for relevant shifts) whereas dephasing is an element of the depolarising process already modelled, and even damping errors are first-order correctable by the extrapolation mitigation that we incorporate. 
}

\change{
For the latter point regarding sampling costs, we emphasise that our numerical studies as reported in Fig.~{\ref{fig:noisy_compile}} correspond to the high-sampling limit, i.e. the performance is that which is achieved when sampling is cheap enough that shot-noise is never the dominant imperfection. Thus the present work does not incorporate optimisations for frugal sampling; indeed exploration of VQAs using this constraint is an evolving theme in its own right (see e.g. Refs.}~\cite{frugal1,frugal2}).
\change{
We can introduce an \textit{ad hoc} shot-noise factor to our model, but because our core gradient estimation does not employ shot-frugal techniques any resulting estimate of absolute sample count required would have little value as an indicator of the performance of a mature, full-optimised application. However, we can obtain some broad indications of the \textit{scaling} of sample cost as we introduce gate noise. To do so, we indeed wrote a function to synthesise shot-noise variability and set a high sample count of $10^8$ for the low-noise ($0.1\%$) limit of the circuit explored in Fig.~{\ref{fig:noisy_compile}}. At such a high sampling rate, there is negligible impact of shot noise versus infinite sampling. We then explore the increase in sampling count needed
to maintain near-optimal performance for gate noise scenarios in the range $0.1\%$ to $1\%$, where by `near-optimal' we mean that the final cost function value should change only modestly (well within a factor of two). We found that a further factor of 3 in the sample count sufficed for per-gate noise of $0.4\%$, while the factor increased to 15 for the high-noise limit of $1\%$. 
 
This simple exercise supports the intuition that it may ultimately be profligate sampling requirements, rather than any inherent limitations of the scheme, that will bound the use of noisy circuits in practical compilation scenarios. 
(As a footnote we remark that for all these simulations the sampling count for the quantities in the metric tensor was set ten times higher, as this object is crucial to correct parameter evolution but, in an optimised setting, is relatively cheap to estimate as it does not need to be updated frequently~{\cite{balintNGcost}}). 
} 

\color{black}  

\section{Numerical scaling}
\label{sec:numerical_scaling}

We now explore the scaling of our recompilation technique as the number of qubits increases. For computational tractability we take the gate-noise-free case and perform only the base recompilation strategy via imaginary time evolution (without additional gate elimination nor luring). However we retain the use of our adaptive timestep scheme as described earlier. 

\begin{figure}
    \centering
    \includegraphics[width=.45\textwidth]{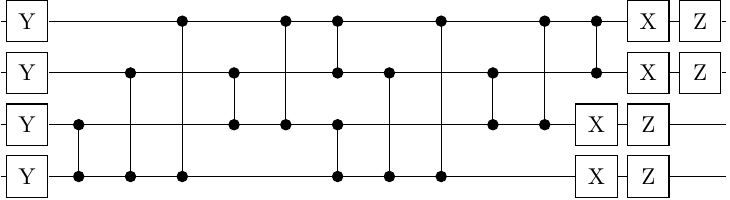}
    \caption{
    An example of circuit $\mathcal{A}$ (Eqn.~{\ref{eq:squeeze_unitary}}) which prepares a $4$-qubit squeezed state{~\cite{squeeze_circuit}} when applied to $\ket{0}$, as used as input to test our recompilation scheme's scaling performance.
    }
    \label{fig:squeeze_circ_example}
\end{figure}

\begin{figure}
    \centering
    \begin{tikzpicture}
    \node[] at (0,1.25) {$\mathcal{A}$};
    \node[inner sep=0pt] (squeezed) at (0,0)
    {\includegraphics[width=.125\textwidth]{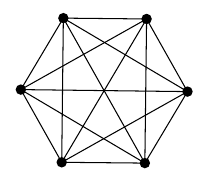}};
    \node[] at (3,1.25) {$\mathcal{B}$};
    \node[inner sep=0pt] (ansatz) at (3,0)
    {\includegraphics[width=.125\textwidth]{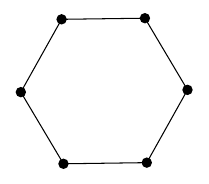}};
    \end{tikzpicture}
    \caption{
    Connectivity of the qubits in the input circuit $\mathcal{A}$ (producing squeezed states), and the target ansatz $\mathcal{B}$, shown here for $6$ qubits.
    }
    \label{fig:scaling_connectivity}
\end{figure}

As input to the recompilation, we choose a scalable family of circuits which generate squeezed states~{\cite{squeezing,squeeze_circuit}}, chiefly
\begin{align}
    \mathcal{A}(\phi_1,\phi_2,\phi_3) 
    = & \prod_t Y_t(\pi/2)
    \prod_{c,t>c} \text{C}_cZ_t(\phi_1) \; \times \nonumber \\
    & \prod_{t,c>t} \text{C}_cZ_t(\phi_1)
    \prod_t X_t(\phi_2)Z_t(\phi_3),
    \label{eq:squeeze_unitary}
\end{align}
where $U_t$ notates gate $U$ acting on qubit $t$, and $C_cU_t$ notates the same gate with additional control qubit $c$. As before, $X,Y,Z$ denote rotations. For each test of a given circuit size, the parameters $\phi_1, \phi_2, \phi_3$ are chosen uniformly randomly in $[0, 2\pi)$. An example circuit which prepares a $4$-qubit squeezed state is shown in
Figure~{\ref{fig:squeeze_circ_example}}, and the qubit connectivity illustrated in Figure~{\ref{fig:scaling_connectivity}}.

\begin{figure}
    \centering
    \includegraphics[width=.49\textwidth]{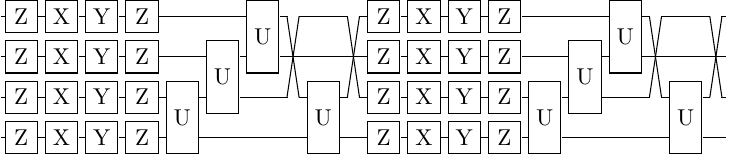}
    \caption{
    A 4-qubit example of $V$ (the first half of  $\mathcal{B}(\vec\theta)$ in Equ~{\ref{eq:squeeze_recomp_template}}), used a template to recompile into in our scaling tests. The second half of $\mathcal{B}$ is merely the reverse of this circuit, with independent parameters.
    }
    \label{fig:b_sq_circ_example}
\end{figure}

With a family of input circuits chosen, we then choose a family of target template circuits with a distinct gate topology and gate set from the input circuit. We nominate the template
\begin{gather}
    \mathcal{B}(\vec\theta) =  V \, \text{reverse}( V), \\
    \text{where}\hspace{.5cm}
    V = \left( \prod_t Z_t X_t Y_t Z_t \prod_t U_{t,t+1} \right)^2, 
    \\
    \text{and}\hspace{.5cm}
    U = \begin{pmatrix} 
    1 & & & \\
    & & e^{i\theta} \\
    & e^{i \theta} & & \\
    & & & 1
    \end{pmatrix},
    \label{eq:squeeze_recomp_template}
\end{gather}
where every gate features a unique parameter $\theta$, and $U$ is the parameterised SWAP gate. An example $4$-qubit $\mathcal{B}(\vec\theta)$ circuit is shown in Figure~{\ref{fig:b_sq_circ_example}}, and the qubit connectivity in Figure~{\ref{fig:scaling_connectivity}}.

Note that $\mathcal{A}(\phi_1,\phi_2,\phi_3)$ and $\mathcal{B(\vec\theta)}$ differ significantly, and there is no exact assignment $\vec\theta$ for which $\mathcal{A}(\phi_1,\phi_2,\phi_3)\ket{0} \equiv \mathcal{B}(\vec\theta)\ket{0}$ besides when both are the identity ($\phi_i = \theta_j = 0$). Furthermore, the \textit{optimum} value of $\vec\theta$ for which $|\bra{0}\mathcal{B}^\dagger(\vec\theta) \mathcal{A}(\phi_1,\phi_2,\phi_3)\ket{0}|$ is maximum, may admit a progressively less-accurate recompilation as the number of qubits increases. Hence, qualifying the performance of the scheme in a scale-invariant way is difficult.

Recompilation will be driven by the simple $n$-qubit Hamiltonian
\begin{align}
    H_\text{rec} = \mathbb{1} - \ket{0}\bra{0}^{\otimes n}.
\end{align}
We opt to use this since its spectral gap remains $1$ as the system size $n$ grows; the first excited state is $(2^n-1)$-degenerate, and the fidelity of the recompiled state is exactly $1 - \braket{E}$, where $\braket{E}$ is the expected energy under $H_\text{rec}$. Furthermore, \change{for the sake of numerical simulation,} it offers quick \change{classical} evaluation ($\mathcal{O}(1)$ and $\mathcal{O}(2^n)$ respectively) of the energy and its partial derivative in parameter space,
\begin{gather}
    \braket{E} = \braket{\psi|H_\text{rec}|\psi} = 1 - |\psi_0|^2,
    \\
    \begin{aligned}
    \nabla \braket{E}_j &= \text{Re}\left\{ 
        \braket{\psi|H_\text{rec}|\frac{\partial \psi}{\partial \theta_j}} \right\}
        \\
        &= \text{Re}\left\{  \braket{\psi|\frac{\partial \psi}{\partial \theta_j}} - \psi_0^* (\frac{\partial \psi}{\partial \theta_j})_0 
        \right\}.
    \end{aligned}
\end{gather}
\change{For a general Hamiltonian, these quantities could otherwise take time $\mathcal{O}(2^{2n})$ to evaluate.}
Here $\psi_0$ indicates the first complex element of the statevector $\ket{\psi}$. These quantities are calculated for every assignment of $\vec\theta$ (i.e. every iteration).

We will pick a number of iterations sufficiently large such that all tests have converged (in our tests, $200$ iterations), and measure the final fidelity achieved. Recompilation will be driven by imaginary time evolution, without luring, though \textit{with} the adaptive time-step subroutine introduced in Section~\ref{sec:adaptive_timestep_intro}.

\begin{figure}
    \centering
    \includegraphics[width=.5\textwidth]{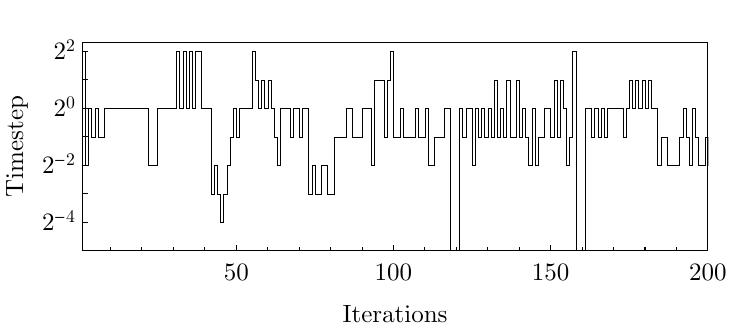}
    \caption{
    An example of how the time-step exponentially evolves under the adaptive scheme used in the scaling tests.
    }
    \label{fig:adaptive_timestep_demo}
\end{figure}

In the subsequent simulations, use of adaptive time-step over a heuristically-motivated fixed time-step improves convergence drastically. The cost is very modest: on average about $1\%$ more energy measurements. A demonstration of the time-step evolving adaptively during one of the subsequent scaling tests is presented in Figure~{\ref{fig:adaptive_timestep_demo}}.

\subsection{Scaling performance}

\begin{figure}
    \centering
    \includegraphics[width=.5\textwidth]{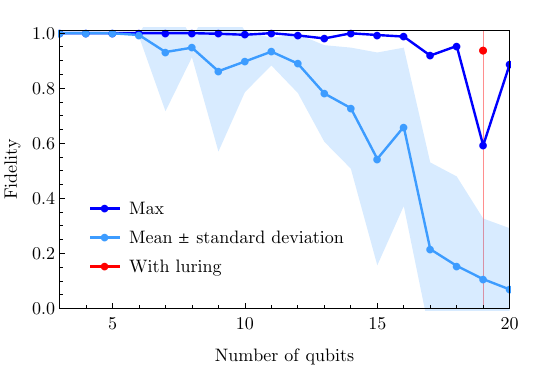}
    \caption{
    Performance of recompilation of squeezed states, driven by $200$ iterations of imaginary time evolution, tested $20$ times for each system size. All tests employ no luring except the red point; this corresponds to the use of $10$ stages of luring achieving a fidelity increase from $59\%$ to $94\%$.
    }
    \label{fig:recompiler_scaling}
\end{figure}
\begin{figure}
    \centering
    \includegraphics[width=.5\textwidth]{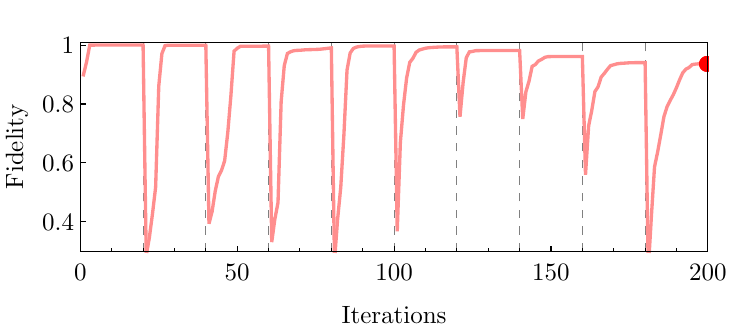}
    \caption{
    Recompilation of the anomalously difficult $19$-qubit squeezed state in Fig~{\ref{fig:recompiler_scaling}}, using $10$ stages of luring. The ultimate fidelity of $94\%$ is denoted in Figure~{\ref{fig:recompiler_scaling}} as a red dot.
    }
   \label{fig:squeezed_luring}
\end{figure}

Using this adaptive time-step scheme, we simulated recompilation of squeezed states of $3$ to $20$ qubits. For each state size, we perform $20$ tests with random squeezed state parameters $\vec\phi$, and random initial ansatz parameters $\vec\theta$. At each of the $200$ iterations of variational imaginary-time evolution, the linear equations were solved using singular value decomposition, with no truncation.
The results are presented in Figure~{\ref{fig:recompiler_scaling}}. 
Despite excellent recompilation of the $16$-qubit state to $99\%$ fidelity, larger squeezed states performed significantly worse. The maximum fidelity achieved for $19$ qubits (highlighted along the red line in Figure~{\ref{fig:recompiler_scaling}}) is $59\%$, though the average is $11\%$.
\change{
We speculate this particular squeezed state may be anomalously difficult to recompile due to, for example, prescribing strong entanglement between non-neighbouring qubits, for which there may be a small space of proximate ansatz states. Converging into such a space from an initial random set of parameters becomes decreasingly likely with increasing number of qubits~{\cite{barren_init}}.
}

As cautioned however, a test of scaling performance may be more difficult than recompiling a particular state. We demonstrate this by recompiling the anomalously difficult $19$ qubit circuit, using the same total number of iterations ($200$), but introducing 10 stages of luring, and parameter truncation. The ansatz parameters now begin at zero, and all parameters in $\mathcal{A}\ket{0}$ (including the $Y(\pi/2)$ terms) are initially scaled to $\times0.1$, and then incrementally restored to their full values over the course of recompilation. We furthermore introduce truncation of singular values below $10^{-5}$ when solving the linear equations involved in variational imaginary-time evolution.
At no extra algorithmic cost, the same $19$ qubit case is now recompiled to a final $94\%$ fidelity, and is denoted by the red point in Figure~{\ref{fig:recompiler_scaling}}. The changing fidelity achieved by recompilation against the lured state is shown in Figure~{\ref{fig:squeezed_luring}}.

\section{Possible applications}

In Section~\ref{sec:numerical_demo}, we demonstrated one possible application area: Using recompilation periodically during some task (modelling dynamics, or eigensolving, etc) so as to `compress' the current circuit and thus `make room' for additional gates. This may be relevant when considerations such as noise accumulation make deeper circuits undesirable. However as a caveat we should stress that in the present demonstration it is actually the recompilation process itself that requires the deepest circuit, since it involves the concatenation $\mathcal{B}^{-1}\mathcal{A}$. A full demonstration of this possibility would require one to show that repeated recompilation can reduce the maximum circuit depth, and this is a topic for further work.

Generally when considering applications it is important to remember that the present technique needs quantum hardware in order to perform the recompilation (unless we are doing so for small circuits relevant to today's small prototypes). Moreover, the recompile process presented here involves a variational eigensolver and therefore will require that circuit $\mathcal{A}$ be executed a large number of times in order to complete the recompile. One might ask, would the user not have done better simply to use the (non-optimal) circuit $\mathcal{A}$ in their application, rather than bothering to recompile it? The answer depends on how many state preparations $\mathcal{A}\ket{\text{in}}$ are required for that application. We now mention two important cases where one might expect that this number is large relative to the recompilation cost, thus making it prudent to indeed recompile and perform subsequent state preparations with $\mathcal{B}\ket{\text{in}}$. 

One such application is that of preparing Gibbs states~\cite{Temme2011,Yung2012}; states of this kind are vital in understanding the equilibrium properties of physical systems and are also relevant to application areas such as machine learning via Boltzmann machines~\cite{Amin2018}. A full understanding of recompilation in the context of Metropolis algorithms would likely require extending the present techniques to include projective measurement, but this does not appear to present an in-principle difficulty. Once one has a recompiled circuit for generation of the Gibbs state, downstream applications involving sampling from that state (typically, a very great number of times) will be correspondingly accelerated. A important context where  repeated sampling of a Gibbs state is essential, is the emerging field of quantum semi-definite programming, see e.g. Refs.~\cite{Brandao2016,vanApeldoorn2018}.

Variational algorithms for the estimation of molecular energies are currently the subject of much interest~\cite{chemReview2018}. Typically some circuit $\mathcal{A}$ of depth $d$ generates an $n$-qubit state that is a good approximation to the ground state of the (translated, qubit-based) chemical \change{Hamiltonian} $H_\text{chem}$~\cite{Troyer2015}. The estimation of $\braket{H_\text{chem}}$ with respect to this state is costly since $H_\text{chem}$ can consist of up to $O(n^4)$ terms~\cite{McClean2016} and the result must be known to at least 3 decimal places to achieve chemical accuracy. Thus the process may take time of order $T\propto dn^4/\epsilon^2$ where $\epsilon$ the tolerable shot noise of order $10^{-3}$ or less. In principle it is then efficient to recompile the circuit prior to the energy estimate, in order to reduce $d$ to $d^\prime$ the depth of the recompiled circuit, provided that (a) the recompile process is rapid compared to $T$ and (b) the circuit recompilation introduces error small compared to $\epsilon$. These criteria can potentially be met since the Hamiltonian within the recompile process itself is very simple, consisting of $n$ terms.

A different type of application would be to use the techniques described to recompile a circuit, initially expressed in some standard language ({\small CNOT}s, single-qubit rotations, etc) into the real gates occurring in some specific device. Such gates will be the device's direct response to control signals (laser or microwave pulses, electrode potential shifts, etc) and may have non-trivial effects not only on targeted qubits but also collateral effects on others, i.e. crosstalk may be significant. To the extent that these effects are deterministic and can be characterised, they can certainly be the building blocks with which we construct our template $\mathcal{B}$. 

This raises an interesting further prospect of what one might call `blind compilation' on a quantum computer:  Provided that the state $\mathcal{A}\ket{\text{in}}$ can be prepared, then the recompilation into template $\mathcal{B}$ can proceed by varying the parameters $\phi_j$ {\it without} understanding the effect of those parameters. We note that the imaginary time eigensolver~\cite{imag2018} which we have applied in the present paper may require modification to be used in this way, but more simple approaches such as gradient descent would be immediately applicable. However the caveat that $\mathcal{A}\ket{\text{in}}$ must be prepared is significant and would seem to imply that some mode of operation of the device does allow the realisation of standard gates (unless $\mathcal{A}\ket{\text{in}}$ has been prepared by another system).

\change{
Note that all these potential applications must carefully consider, and may be ultimately limited by, the total samplings costs. Since a practical implementation must make use of strategies to budget and optimise sampling, the final sampling costs cannot be quickly estimated here. We do at least know that the cost of the metric tensors involved in imaginary time evolution and noisy quantum natural gradient are marginal~{\cite{noisy_natural}}.
}

\section{Conclusion}

We have described a method for recompiling a quantum circuit $\mathcal{A}$ into a new, user-specified `template' with the goal that the resulting circuit $\mathcal{B}$ has the same effect on {\it a specific} input state $\ket{\text{in}}$, i.e. we aim to achieve $\mathcal{A}\ket{\text{in}}=\mathcal{B}\ket{\text{in}}$. This is a quite different (and more permissive) requirement than seeking a new circuit that represents the same overall unitary. However our criterion is exactly the relevant one for many quantum algorithms currently under study. An important class is that of variational hybrid algorithms, whether intended for eigensolving, or modelling the dynamics of quantum systems, or materials science applications. 

The methods we describe include the recompilation process itself, which is achieved through an isomorphism to a ground-state finding protocol (i.e. an eigensolver) and is therefore a hybrid algorithm itself. We  identify an approach which can improve the performance of the recompilation by using a `lure', i.e. providing a proximal goal and then moving the goal further off once it has been approached (this is discussed fully in Appendix~\ref{app:lure}). Furthermore we have described a post-processing gate elimination routine, which can be applied following the recompilation in order to remove `weak' gates at relatively little cost in the fidelity of the circuit. 

We applied these ideas to a specific, complicated $7$-qubit circuit $\mathcal{A}$, which we created by considering a simulation task involving $7$ spins. The creation of the example circuit was itself an interesting task, and we make a number of remarks about it in the appendices. The circuit involved $186$ unique non-Clifford gates and was therefore quite substantial. Our recompilation process successfully realised an equivalent smaller circuit with half the number of two-qubit gates and a smaller number of gate types; the new circuit replicated the action of the original with a $0.998$ fidelity.

Applying the optional post-compile gate elimination, we removed a further thirty gates and reduced the two-qubit gate count to only about a third of the number in original circuit $\mathcal{A}$. This came at the cost of a small further reduction in fidelity, to $0.995$. 

\change{Given that circuit recompilation is likely to be of value in the NISQ era, it is very relevant to ask whether the methods described here can be successful even when the quantum machine employed for the optimisation is imperfect. To explore this question we repeated our initial study but with the introduction of noise throughout the circuits involved. We confirmed a remarkable degree of robustness in the compilation process, concluding that \textit{for any level of noise for which the original circuit can be of practical use}, the imperfect recompilation process would introduce very little further degradation. This observation was made with caveats regarding noise model and sample costs, as explained in the main text. We note that a related but distinct question, is whether compilation (or `learning' of tasks) can discover a more noise-resilient way to approximately perform a task under a given noise model; there have been encouraging results reported on this theme quite recently~{\cite{PRXQuantum2}}. 
}

Having thus explored the $7$-qubit case in some detail, we then studied the scaling of the recompilation protocol over a range of tasks from $3$ to $20$ qubits. Such a study inevitably faces a number of difficulties in obtaining a clear conclusion. In particular, because we wish to see the scaling of a challenging task (recompilation of a meaningful circuit into a `template' with a less rich gate topology) it follows that the task itself may become harder with increasing number of qubits -- then, even an ideal recompilation system would return circuits of diminishing fidelity. We observed that the protocol successfully achieved recompilation at all scales tested, although there was an appreciable drop-off in fidelity for the largest cases. An interesting observation was that the case with worst performance, a certain randomly-generated task involving $19$ qubits, could be dramatically improved by altering the protocol to enable the `lure' technique. A reasonable inference is that, as the size of the recompilation task increases, na\"ive implementations of the variational technique must be replaced with more nuanced protocols.

Ultimately the recompilation method we describe here, being dependent on variational energy-minimisation, will be subject to essentially the same scaling as all variational methods. The corresponding issues of `barren plateaus'~{\cite{barren_init}} and local versus global cost functions~{\cite{cerezoNatComms2021}}, etc, are under very active investigation in the field~
\cite{Yuan2019,natural_gradient,noisy_natural}.

We concluded the main part of the paper by indicating some potential application areas for the technique.

\bigskip
\noindent{\textbf{Acknowledgements}
We thank Xiao Yuan, Sam McArdle and Earl Campbell for very helpful comments. This work was supported by EPSRC grant EP/M013243/1. SCB acknowledges support from the EPSRC QCS Hub EP/T001062/1, from U.S. Army Research Office Grant No. W911NF-16-1-0070 (LOGIQ), and from EU H2020-FETFLAG-03-2018 under the grant agreement No 820495 (AQTION). The authors would like to acknowledge the use of the University of Oxford Advanced Research Computing (ARC) facility in carrying out this work (\href{http://dx.doi.org/10.5281/zenodo.22558}{dx.doi.org/10.5281/zenodo.22558}).
}


\onecolumngrid
\appendix

\section{Demonstration of the {\em Lure} method \label{app:lure}}



\begin{figure}[bt]
    \centering
    \begin{minipage}{.49\textwidth}
    \includegraphics[width=\textwidth]{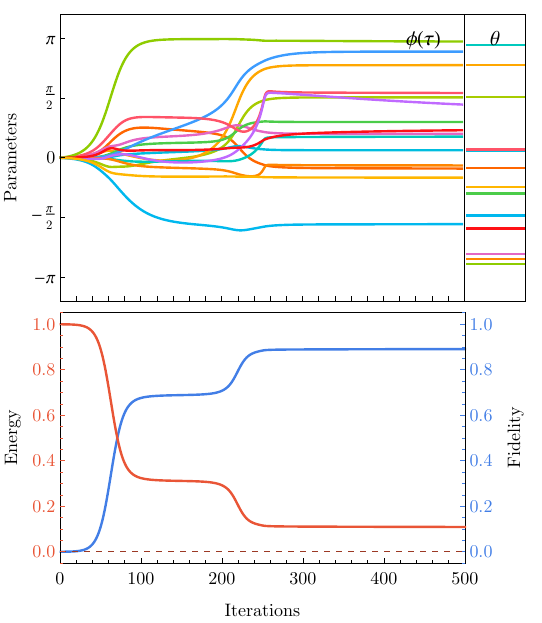}
    \caption{
        A failed direct recompilation of the 5-qubit circuit $\mathcal{C}(\vec{\theta})\ket{0}$ (Eqn.~\ref{eq:example_5qb_ansatz}) into $\mathcal{C}(\vec{\phi})\ket{0}$.
        The top panel shows the 15 parameters converge to a non-ground state, distinct from their optimal (original) parameters on the right. The bottom plot shows the energy of $\mathcal{C}(\vec{\phi})^{-1} \mathcal{C}(\vec{\theta}) \ket{0}$ and fidelity
        $F(\mathcal{C}(\vec{\phi})\ket{0}, \, \mathcal{C}(\vec{\theta})\ket{0})$
        plateau without reaching their optima.
    }
    \label{fig:example_5qb_no_luring}
    \end{minipage} \hfill
    \begin{minipage}{.49\textwidth}
    \includegraphics[width=\textwidth]{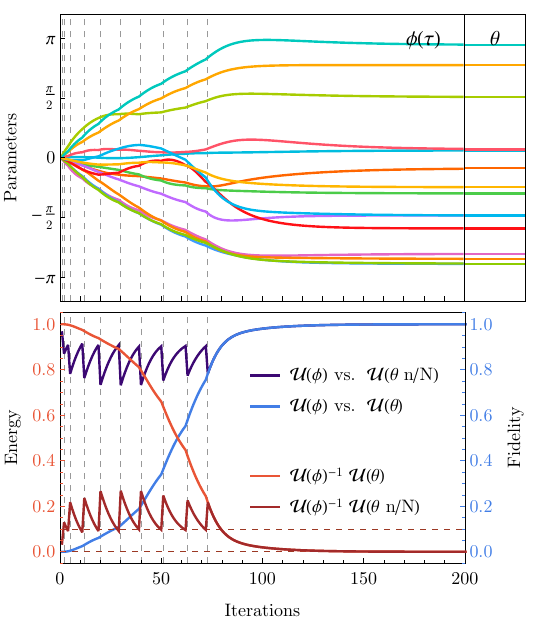}
    \caption{
        A successful recompilation of the 5-qubit ansatz $\mathcal{C}(\vec{\theta})\ket{0}$ (Eqn.~\ref{eq:example_5qb_ansatz}) into $\mathcal{C}(\vec{\phi})\ket{0}$ using \textit{luring}: successively recompiling $\mathcal{C}(\vec{\theta}\,n/10) \ket{0}$ where $n$ is incremented each time the measurable energy falls within $0.1$ of ground (indicated by the dashed lines).
        The top panel shows the evolving parameters $\vec{\phi}$ recover their original counterparts. The bottom panel shows the fidelity and energy reach their optima.
    }
    \label{fig:example_5qb_with_luring}
\end{minipage}
\end{figure}


To illustrate the Lure method, we here consider a 5-qubit 15-gate circuit
\begin{equation}
\mathcal{C}(\vec{\theta}) = \prod\limits_{q=0}^4 
\Rx_q(\theta_{5q}) \Rz_q(\theta_{5q+1}) C_q \Ry_{(q + 1) \% 5}(\theta_{5q + 2}),
\label{eq:example_5qb_ansatz}
\end{equation}
where $Z_q(\theta_j) = \exp(-i \theta_j \sigma^z/2 )$ indicates a rotation of qubit $q$ (indexed from 0) around the $z$-axis of the Bloch sphere by angle $\theta_j$ (and similarly for $Y$ and $Z$), $C_q$ indicates the proceeding gate is controlled on qubit $q$, and $\%$ indicates modulus.
We randomly assign $\vec{\theta}$ and test recompilation of $\mathcal{C}(\vec{\theta}) \ket{0}$ into the same circuit, $\mathcal{C}(\vec{\phi}) \ket{0}$, which can in theory be done with perfect fidelity via $\vec{\phi} = \vec{\theta}$. 
We use time-step $\Delta \tau = 0.1$ and employ a simple Hamiltonian $\hat{H}_{\text{rec}} = \mathbb{1} - \ket{0}\bra{0}$
which under ideal imaginary time evolution, drives $\mathcal{C}(\vec{\phi})^{-1} \mathcal{C}(\vec{\theta}) \ket{0} \rightarrow \ket{0}$. 
Despite this, attempts to directly recompile $\mathcal{C}(\vec{\theta})$ can fail as shown in Fig.~\ref{fig:example_5qb_no_luring}, where the parameters become trapped in a non-ground state.
To combat this, Fig.~\ref{fig:example_5qb_with_luring} demonstrates \textit{luring} by successively
recompiling an intermediate state $\mathcal{C}(\vec{\theta}\, n/10)$ for $n=1,2,\dots10$, updated whenever the energy falls within $0.1$ of the known $0$ energy ground state. With luring, the evolved parameters $\vec{\phi}$ are seen to restore their $\theta$ counterparts.

\section{Li's Algorithm and Trotter comparison\label{app:li_algorithm}}

\begin{figure}[t]
    \centering
    \includegraphics[width=.5\columnwidth]{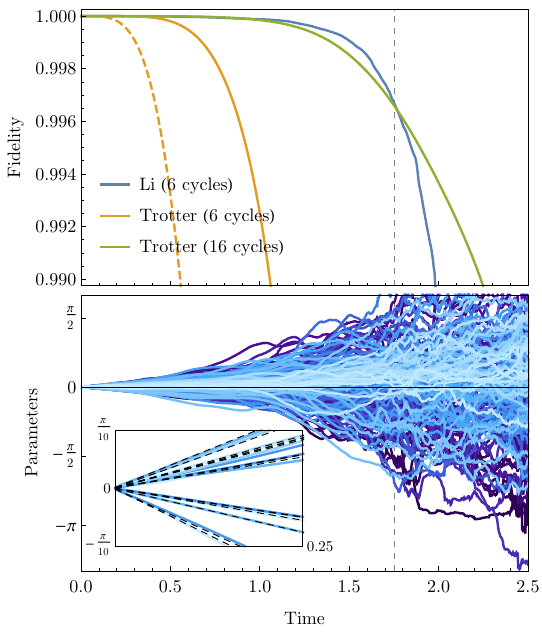}
    \caption{Comparative performance of the Li and Trotter methods for realtime simulation of the spin system specified in Eqn.~(\ref{eqn:spinH}), without any compilation.
    The top panel shows the fidelity with true evolution. The Li method freely evolves the parameters in an ansatz with the structure of 6 Trotter cycles. Meanwhile, the Trotter method uses fixed parameters and every second cycle is reversed except in the simulation indicated by the orange dashed line; this is discussed further in the main text.
    The bottom panel shows how the 186 parameters in Li's method evolve. The subplot compares 16 of the 32 parameters (summed over Trotter cycles) to their fixed Trotter counterparts in the early stages of the simulation.
    The vertical dashed line in both panels indicates the time $t=1.75$ time units at which simulation is stopped and the Li circuits recompiled according to the main text.
    }
    \label{fig:Trotter_comparison}
\end{figure}

\begin{figure}[t]
    \centering
    \includegraphics[width=.5\columnwidth]{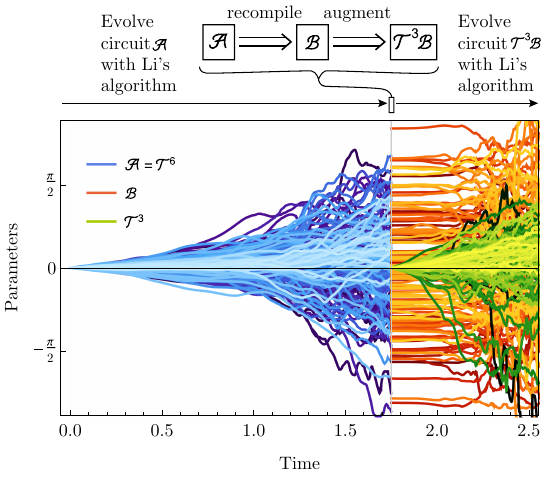}
    \caption{
    A demonstration of continuing realtime simulation by recompiling ansatz $\mathcal{A}(\vec{\theta})$ into $\mathcal{B}(\vec{\phi})$ and appending 3 Trotter cycles with freely evolving parameters $\vec{\omega}$.
    The blue lines indicate $\vec{\theta}$ as in Fig.~\ref{fig:Trotter_comparison} until $t=1.75$.
    Thereafter, the red lines indicate $\vec{\phi}$ and the green lines, $\vec{\omega}$, which are initially zero.
    }
    \label{fig:params_during_extended_realsim}
\end{figure}

In the main paper we generated our input circuit $\mathcal{A}$ by considering a simulation task. Rather than using a direct Trotter method we instead used a gate layout suitable for Trotter but selected the gate parameters using a more efficient method. That method is the one described in Ref.~\cite{YingPRX} by Li and Benjamin, and we refer to it as Li's Algorithm. 

Two variants are described in Ref.~\cite{YingPRX}, each resulting from a different initial variational equation. In the present paper we use the method that results from McLachlan's variational principle, which in practice means selecting $\eta=-i$ in Eqns.~(5) and (6) of that paper, which we reproduce below:
\begin{equation}
    \sum_qM_{k,q}{\dot \lambda}_q=V_k,
    \label{eqn:Li}
\end{equation}
where $\lambda_q$ are the parameters in the evolving circuit, equivalent to the $\theta_i$ or $\phi_i$ in our treatments. Meanwhile 
\begin{eqnarray}
M_{k,q}&=&\frac{\partial\bra{\Psi}}{\partial\lambda_k}
\frac{\partial\ket{\Psi}}{\partial\lambda_q}+\text{H.C.},
\label{eqn:matrixM}
\\
V_k&=&-i\frac{\partial\bra{\Psi}}{\partial\lambda_k}H
\ket{\Psi}+\text{H.C.}
\label{eqn:vectorV}
\end{eqnarray}
In Ref.~\cite{YingPRX} it is shown that these matrix and vector elements are of the form $a\text{Re}(e^{i\theta}\braket{\bar 0\vert U \vert \bar 0})$ where $U$ is a unitary involving the $k^\text{th}$ and $q^\text{th}$ gates of the parameterised circuit, and $\ket{\bar 0}$ is the all-zero state. 

It is further shown that $\text{Re}(e^{i\theta}\braket{\bar 0\vert U \vert \bar 0})=\text{Tr}(X_\text{anc}\rho)$, i.e. it relates directly to expected value of an ancilla that probes the parameterised circuit in a simple fashion (see Fig.~2 of Ref.~\cite{YingPRX}, which is a variant of a circuit proposed in Ref.~\cite{Ekert2002} in 2002).

Therefore a quantum computer is able to provide the $M$ and $V$ values involved in equation (\ref{eqn:Li}), which can then be solved classically to yield the $\dot\lambda_i$ values and thus to update the parameters for the next incremental state of the circuit. 

In the present work, a small modification was applied to the algorithm specified in Ref.~\cite{YingPRX}. We now ensure there is a parameter in the problem which controls the global phase. This can be significant since the McLachlan principle is sensitive to global phase, even though it has no physical meaning. Generally one need not introduce an actual global phase gate, instead such a gate can be virtually present and allowed for in the update equations. Fortunately in the specific case we describe in the main paper, the input state is such that a gate performing a $z$-rotation gate acts on a qubit in state $\ket{1}$, thus this gate indeed generates a global phase.

In creating our trial circuit $\mathcal{A}$, we made some interesting observations concerning the way in which the circuit derived using Li's algorithm compares to a standard Trotter circuit. A comparison of the Li and Trotter methods is shown in Fig.~\ref{fig:Trotter_comparison}, which includes a visualisation of the Li parameters in the bottom panel. 
By `standard Trotter' we mean that the choice of parameters (angles $\theta$ within the gates) is determined simply from the elapsed simulation time and the strength of the corresponding Hamiltonian term.

Figure~\ref{fig:Trotter_comparison} includes three lines (orange-dashed, orange and green) which correspond to Trotter-type solutions of this kind. The orange-dashed line is a direct realisation of the circuit pattern shown in Fig.~\ref{fig:circuitA}. Its performance is poor since the simple repetition of identical cycles causes Trotter errors to directly accumulate. The orange line represents the best realisation of the Trotter-type approach that we found: it involves alternatingly inverting the order of the gates (versus that shown in the middle panel of Fig.~\ref{fig:circuitA}). This is related to approaches in Ref.~\cite{Childs2018}, but without the randomisation element; in our simulations random re-ordering was not advantageous, but this may relate to the small system size and circuit depth. We acknowledge that there are many approaches including the sampling method of Campbell which has just been presented~\cite{Campbell2018}, and we have not comprehensively searched over them. In short, the orange line is a fair indicator of a `smart' Trotter approach (as opposed to the naive one represented by the orange-dashed line). The green line is the same approach of alternating inversions, but now with a total of $16$ cycles. 

The figure indicates that Li's algorithm does better than a (smart, if not necessarily optimal) Trotter approach for a given circuit depth. This is to be expected since Li's approach has greater freedom: it does not seek to approximate the time evolution unitary but rather it seeks the circuit with the closest-to-ideal action on the defined input state.

As an aside, we remark that since the ansatz has the structure of 6 Trotter cycles, it is natural to wonder whether Li's algorithm will, at early times, choose the same set of parameters which Trotterisation prescribes to the rotation gates in the time-evolution unitary for those times. Indeed, around half of Li's parameters (when summed according to their Trotter term) precisely follow their Trotter counterparts, as shown in the subfigure of Fig.~\ref{fig:Trotter_comparison}. Interestingly, the other parameters which differ do improve the fidelity achieved by Li's algorithm versus Trotterisation by an appreciable factor even in the early evolution (the fidelity `gap' is $4 \times 10^{-4}$ after 100 iterations of $\Delta t = 2.5 \times 10^{-3}$).

This behaviour is even seen in the parameters of the additional Trotter cycles appended to the recompiled ansatz, as shown in Fig.~\ref{fig:params_during_extended_realsim}.

\section{The imaginary time eigensolver\label{app:imag-time}}
The recompilation process described in the main paper employs a ground state finding algorithm at its core. We seek the ground state of a fictitious Hamiltonian $H_\text{rec}$ because by doing so we necessarily find a circuit that can invert the original circuit $\mathcal{A}$, and thus we obtain an alternative realisation of $\mathcal{A}$.

There are a variety of possible algorithms that can configure a parameterised circuit so that it maps a fixed input state onto (an approximation to) the ground state of a specified Hamiltonian. In the present paper we employed a recent imaginary time method, as described in Ref.~\cite{imag2018}. In essence this algorithm is a variant of Li's algorithm described in Appendix~\ref{app:li_algorithm} above; as proved in Ref.~\cite{imag2018} we need only modify Eqn.~(\ref{eqn:vectorV}) with a factor $i$ in order to switch from a real time evolution (i.e. an evolution according to the Schr\"odinger equation) to an evolution under $\exp(-H\,t)$ (with renormalisation) which drives the system to its ground state -- or rather, to the lowest eigenstate with finite projection on the initial state.

\section{Details of numerical simulation\label{app:numerical_sim_details}}

We simulate the quantum circuits involved in the introduced technique using the Quantum Exact Simulation Toolkit~\cite{quest18}, 
\change{ 
and QuESTlink~{\cite{questlink19}}.
}
\change{ 
All numerics were performed using GPU parallelisation, on a desktop machine with a 24\,GiB Quadro P6000 GPU.
}

Several computational shortcuts are taken in the evaluation of the equations involved in the Li method, which we stipulate in the Supplementary Materials of previous works~\cite{imag2018,suguru18}. 
Tables~\ref{tab:hsys_coeffs} and \ref{tab:state_before_recomp} list some numerical values used in the simulated Hamiltonian and recompilation process.
The full simulation code for the demonstration in Sec.~\ref{sec:numerical_demo}, which depends only on the GNU Scientific Library~\cite{GslGnuNumericsLibrary} and standard/included C libraries, is available on GitHub~\cite{githubLink}. 
\change{
This includes extensive documentation about employing the recompiler as a simulation tool.
The QuESTlink and Mathematica code for the scaling tests in Sec.~{\ref{sec:numerical_scaling}} are also available on GitHub~{\cite{mmaGithubLink}}
}

The Li method, in both its original form~\cite{YingPRX} and imaginary-time adaptation~\cite{imag2018}, involves populating a family of linear equations which are then numerically solved by a classical machine.
We make some observations about the choice of the linear solving algorithm, which was previously seen to affect the ground state convergence rate of imaginary-time simulations~\cite{suguru18}. In this work, we discover that Tikhonov regularisation and least squares minimisation~\cite{gslDiagAlgorithms} work well to ensure the parameters vary smoothly in realtime Li simulation. However, imaginary time Li simulation benefited from truncated singular value decomposition (TSVD)~\cite{gslDiagAlgorithms} which imposes no constraint on the parameter smoothness and saw much faster convergence to the ground state than the previously mentioned methods.


\change{
We now outline the approach taken for tractable simulation of the fully noise-burdened circuits in Section~{\ref{sec:robustness}}. A lab-based realisation of the method used here, specifically the experimental determination of the crucial metric tensor employed in the imaginary time (or `natural gradient') method, would require two copies of the main register in order to perform swap-test-like process as explained in Ref.~{\cite{noisy_natural}}. However, in the present case of numerical simulation, we can benefit from the fact that the density matrix corresponding to each register is identical (presuming the same noise model through the device). We can avoid the very considerable increase in simulation costs that would arise from explicitly representing the density matrix of $1+2n$ qubits, instead representing $1+n$. This is exact until the actual swap-test process; the noise afflicting this stage (which is far more shallow than the preceding ansatz-based circuits) can be modelled by estimating how severely the observables comprising  the metric tensor would be impacted, and then indeed applying this level of corruption directly to the tensor. In our study we `err on the side of caution' by making these noise events quite severe, in order to be certain that the imperfection of the swap-test phase does not undermine the robustness of the overall compilation process. Specifically: we first implement a proxy for the inter-register swap-test via a control-permute of all qubits in the (sole) register, such that every qubit position is changed (formally, a derangement). We then characterise the degree of corruption in a random sample of $30$ single- and two-qubit Pauli observables (themselves computed using noisy gates). The noise-free metric tensor is then corrupted by corresponding variances. 
}

\change{
As a side note: We employed the following heuristic rule, entirely compatible with experimental realisation, which somewhat improved performance in the high-noise limit. Specifically, when taking a step in the optimisation walk, the protocol tries both the `new' metric tensor and the `old' tensor, where the latter was the last successfully computed tensor. If the old one gives rise to a superior step down in cost function, then this is employed instead of the new (which is deemed to be too corrupted to be useful). This approach exploits the fact that the metric tensor should be smoothly changing function of the ansatz parameters, therefore one can `fall back' to an immediately prior tensor without significant impact on efficacy. 
}

Finally, we now list the chosen numerical constants.
For realtime and imaginary time simulation, we use step sizes of $\Delta t = 2.5 \times 10^{-3}$ and $\Delta \tau = 1 \times 10^{-2}$ respectively. We regularise the change in parameters per-iteration using truncated singular value decomposition (TSVD) and Tikhonov regularisation. We use a TSVD tolerance of $10^{-5}$ and choose the Tikhonov parameter as the corner of a 3-point L-curve~\cite{gslDiagAlgorithms}. Wavefunction derivatives are estimated using a fourth-order finite difference method with a step-size of $\Delta \theta = 10^{-5}$. When we desire an initially identity unitary, we set all parameters $\theta_j=10^{-8}$ to avoid singularities in the matrix inversion during the first iteration. During the gate elimination subroutine, we restrict the to-be-zeroed parameter to vary by at most $|\Delta \phi_j| \le 0.1$ radians each iteration.

\begin{table}
\centering
\begin{tabular}[t]{|c|c|}
\hline $B_{1}$ & -0.433333 \\
\hline $B_{2}$ & -1.006667 \\
\hline $B_{3}$ & -0.941460 \\
\hline $B_{4}$ & -0.312000 \\
\hline $B_{5}$ & -0.478667 \\
\hline $B_{6}$ & -0.347867 \\
\hline $B_{7}$ & -0.314533 \\
\hline
\end{tabular}
\quad
\begin{tabular}[t]{|c|c|}
\hline $J_{1,2}^{x}$ & 0.730767 \\
\hline $J_{2,3}^{x}$ & 0.745333 \\
\hline $J_{1,3}^{x}$ & 0.830400 \\
\hline $J_{3,4}^{x}$ & 0.521333 \\
\hline $J_{4,5}^{x}$ & 0.543800 \\
\hline $J_{5,6}^{x}$ & 0.338771 \\
\hline $J_{6,7}^{x}$ & 0.700544 \\
\hline $J_{7,5}^{x}$ & 0.771333 \\
\hline $J_{1,2}^{y}$ & 0.367333 \\
\hline $J_{2,3}^{y}$ & 0.913333 \\
\hline $J_{1,3}^{y}$ & 0.570667 \\
\hline $J_{3,4}^{y}$ & 0.315333 \\
\hline
\end{tabular}
\quad
\begin{tabular}[t]{|c|c|}
\hline $J_{4,5}^{y}$ & 0.412000 \\
\hline $J_{5,6}^{y}$ & 0.641200 \\
\hline $J_{6,7}^{y}$ & 0.697453 \\
\hline $J_{7,5}^{y}$ & 0.345333 \\
\hline $J_{1,2}^{z}$ & 0.362667 \\
\hline $J_{2,3}^{z}$ & 0.864433 \\
\hline $J_{1,3}^{z}$ & 0.527000 \\
\hline $J_{3,4}^{z}$ & 0.543000 \\
\hline $J_{4,5}^{z}$ & 0.547667 \\
\hline $J_{5,6}^{z}$ & 1.014333 \\
\hline $J_{6,7}^{z}$ & 0.368100 \\
\hline $J_{7,5}^{z}$ & 0.701000 \\
\hline
\end{tabular}
\caption{The coefficients of $\hat{H}_{\text{sys}}$, the 7-qubit spin network Hamiltonian. These are available in plaintext in the online repository~\cite{githubLink} (\href{https://github.com/QTechTheory/DissipativeRecompiler/blob/master/data/hamil_sim.txt}{direct link}).}
\label{tab:hsys_coeffs}
\end{table}

{
\begin{table}
\centering
\resizebox{\textwidth}{!}{%
\begin{tabular}{|c|c|c|c|c|c|c|c|c|c|c|c|c|c|c|c|c|c|}
\cline{1-3}
n & $\hat{O}_n$ & $\theta_n$ 
\\
\hline \textbf{1} & $\Rz_0$ & -0.00078  & \textbf{32} & $\Rz_0$ & 0.00612  &  \textbf{63} & $\Rz_0$ & -1.92631  &  \textbf{94} & $\Rz_0$ & 0.07784  &  \textbf{125} & $\Rz_0$ & 0.26960 &  \textbf{156} & $\Rz_0$ & -0.80412 \\
\hline \textbf{2} & $\Rz_1$ & -0.76524  & \textbf{33} & $\Rz_1$ & -0.72202  &  \textbf{64} & $\Rz_1$ & -0.72235  &  \textbf{95} & $\Rz_1$ & -0.28875  &  \textbf{126} & $\Rz_1$ & -0.14207 &  \textbf{157} & $\Rz_1$ & -1.95611 \\
\hline \textbf{3} & $\Rz_2$ & -0.07102  & \textbf{34} & $\Rz_2$ & -0.89156  &  \textbf{65} & $\Rz_2$ & -0.98678  &  \textbf{96} & $\Rz_2$ & 0.30971  &  \textbf{127} & $\Rz_2$ & 0.06518 &  \textbf{158} & $\Rz_2$ & -1.35826 \\
\hline \textbf{4} & $\Rz_3$ & -1.57210  & \textbf{35} & $\Rz_3$ & 1.92348  &  \textbf{66} & $\Rz_3$ & 0.34984  &  \textbf{97} & $\Rz_3$ & -1.02248  &  \textbf{128} & $\Rz_3$ & 0.10542 &  \textbf{159} & $\Rz_3$ & -0.83901 \\
\hline \textbf{5} & $\Rz_4$ & -0.57014  & \textbf{36} & $\Rz_4$ & -0.31031  &  \textbf{67} & $\Rz_4$ & -0.42137  &  \textbf{98} & $\Rz_4$ & 0.08347  &  \textbf{129} & $\Rz_4$ & -0.61129 &  \textbf{160} & $\Rz_4$ & -0.21782 \\
\hline \textbf{6} & $\Rz_5$ & -0.03702  & \textbf{37} & $\Rz_5$ & -0.54383  &  \textbf{68} & $\Rz_5$ & 0.20952  &  \textbf{99} & $\Rz_5$ & -0.68914  &  \textbf{130} & $\Rz_5$ & 1.16104 &  \textbf{161} & $\Rz_5$ & -0.85344 \\
\hline \textbf{7} & $\Rz_6$ & -0.56917  & \textbf{38} & $\Rz_6$ & 0.42303  &  \textbf{69} & $\Rz_6$ & -0.64586  &  \textbf{100} & $\Rz_6$ & -2.11670  &  \textbf{131} & $\Rz_6$ & -0.49693 &  \textbf{162} & $\Rz_6$ & -0.36461 \\
\hline \textbf{8} & $\Rx_{0,1}$ & -0.23393  & \textbf{39} & $\Rx_{0,1}$ & -0.23944  &  \textbf{70} & $\Rx_{0,1}$ & -1.19517  &  \textbf{101} & $\Rx_{0,1}$ & 0.93936  &  \textbf{132} & $\Rx_{0,1}$ & 0.98615 &  \textbf{163} & $\Rx_{0,1}$ & 1.56663 \\
\hline \textbf{9} & $\Rx_{1,2}$ & -0.77487  & \textbf{40} & $\Rx_{1,2}$ & 1.10291  &  \textbf{71} & $\Rx_{1,2}$ & 0.19727  &  \textbf{102} & $\Rx_{1,2}$ & 0.99647  &  \textbf{133} & $\Rx_{1,2}$ & 0.31287 &  \textbf{164} & $\Rx_{1,2}$ & 0.67684 \\
\hline \textbf{10} & $\Rx_{0,2}$ & 0.87379  & \textbf{41} & $\Rx_{0,2}$ & 0.62165  &  \textbf{72} & $\Rx_{0,2}$ & 0.43687  &  \textbf{103} & $\Rx_{0,2}$ & 0.02575  &  \textbf{134} & $\Rx_{0,2}$ & 1.18330 &  \textbf{165} & $\Rx_{0,2}$ & 0.73790 \\
\hline \textbf{11} & $\Rx_{2,3}$ & -1.09072  & \textbf{42} & $\Rx_{2,3}$ & 0.22439  &  \textbf{73} & $\Rx_{2,3}$ & 0.77102  &  \textbf{104} & $\Rx_{2,3}$ & 0.90254  &  \textbf{135} & $\Rx_{2,3}$ & 0.24543 &  \textbf{166} & $\Rx_{2,3}$ & 0.29939 \\
\hline \textbf{12} & $\Rx_{3,4}$ & 0.81605  & \textbf{43} & $\Rx_{3,4}$ & 1.58883  &  \textbf{74} & $\Rx_{3,4}$ & -0.08047  &  \textbf{105} & $\Rx_{3,4}$ & -0.33305  &  \textbf{136} & $\Rx_{3,4}$ & 0.50658 &  \textbf{167} & $\Rx_{3,4}$ & 0.10799 \\
\hline \textbf{13} & $\Rx_{4,5}$ & -0.10714  & \textbf{44} & $\Rx_{4,5}$ & 0.69977  &  \textbf{75} & $\Rx_{4,5}$ & -0.11872  &  \textbf{106} & $\Rx_{4,5}$ & 0.04207  &  \textbf{137} & $\Rx_{4,5}$ & 0.50517 &  \textbf{168} & $\Rx_{4,5}$ & 0.31331 \\
\hline \textbf{14} & $\Rx_{5,6}$ & -1.61988  & \textbf{45} & $\Rx_{5,6}$ & -0.43299  &  \textbf{76} & $\Rx_{5,6}$ & 1.13964  &  \textbf{107} & $\Rx_{5,6}$ & 1.74800  &  \textbf{138} & $\Rx_{5,6}$ & -1.30005 &  \textbf{169} & $\Rx_{5,6}$ & 0.48987 \\
\hline \textbf{15} & $\Rx_{6,4}$ & -0.58697  & \textbf{46} & $\Rx_{6,4}$ & 0.38593  &  \textbf{77} & $\Rx_{6,4}$ & 0.26760  &  \textbf{108} & $\Rx_{6,4}$ & 0.19480  &  \textbf{139} & $\Rx_{6,4}$ & 0.18067 &  \textbf{170} & $\Rx_{6,4}$ & 0.38603 \\
\hline \textbf{16} & $\Ry_{0,1}$ & 1.03841  & \textbf{47} & $\Ry_{0,1}$ & 0.28201  &  \textbf{78} & $\Ry_{0,1}$ & 0.03540  &  \textbf{109} & $\Ry_{0,1}$ & 0.64047  &  \textbf{140} & $\Ry_{0,1}$ & -0.98891 &  \textbf{171} & $\Ry_{0,1}$ & 0.59770 \\
\hline \textbf{17} & $\Ry_{1,2}$ & -0.20021  & \textbf{48} & $\Ry_{1,2}$ & 0.77338  &  \textbf{79} & $\Ry_{1,2}$ & 0.40912  &  \textbf{110} & $\Ry_{1,2}$ & 0.75941  &  \textbf{141} & $\Ry_{1,2}$ & 0.78280 &  \textbf{172} & $\Ry_{1,2}$ & 0.89393 \\
\hline \textbf{18} & $\Ry_{0,2}$ & 1.25312  & \textbf{49} & $\Ry_{0,2}$ & 0.17709  &  \textbf{80} & $\Ry_{0,2}$ & 0.25703  &  \textbf{111} & $\Ry_{0,2}$ & 0.04256  &  \textbf{142} & $\Ry_{0,2}$ & 0.62293 &  \textbf{173} & $\Ry_{0,2}$ & 0.22623 \\
\hline \textbf{19} & $\Ry_{2,3}$ & 0.19017  & \textbf{50} & $\Ry_{2,3}$ & 0.37472  &  \textbf{81} & $\Ry_{2,3}$ & -0.00260  &  \textbf{112} & $\Ry_{2,3}$ & 0.43709  &  \textbf{143} & $\Ry_{2,3}$ & 0.25405 &  \textbf{174} & $\Ry_{2,3}$ & 0.35892 \\
\hline \textbf{20} & $\Ry_{3,4}$ & 0.59210  & \textbf{51} & $\Ry_{3,4}$ & 0.59450  &  \textbf{82} & $\Ry_{3,4}$ & -0.06945  &  \textbf{113} & $\Ry_{3,4}$ & 0.89857  &  \textbf{144} & $\Ry_{3,4}$ & 0.03360 &  \textbf{175} & $\Ry_{3,4}$ & 0.25520 \\
\hline \textbf{21} & $\Ry_{4,5}$ & -1.71059  & \textbf{52} & $\Ry_{4,5}$ & -0.22396  &  \textbf{83} & $\Ry_{4,5}$ & 0.25205  &  \textbf{114} & $\Ry_{4,5}$ & 0.28955  &  \textbf{145} & $\Ry_{4,5}$ & 0.42917 &  \textbf{176} & $\Ry_{4,5}$ & 0.35523 \\
\hline \textbf{22} & $\Ry_{5,6}$ & 0.05980  & \textbf{53} & $\Ry_{5,6}$ & -0.02429  &  \textbf{84} & $\Ry_{5,6}$ & -0.44813  &  \textbf{115} & $\Ry_{5,6}$ & 2.37157  &  \textbf{146} & $\Ry_{5,6}$ & -0.84443 &  \textbf{177} & $\Ry_{5,6}$ & 0.88797 \\
\hline \textbf{23} & $\Ry_{6,4}$ & -0.12607  & \textbf{54} & $\Ry_{6,4}$ & -0.23800  &  \textbf{85} & $\Ry_{6,4}$ & 0.38207  &  \textbf{116} & $\Ry_{6,4}$ & 0.29358  &  \textbf{147} & $\Ry_{6,4}$ & 0.43788 &  \textbf{178} & $\Ry_{6,4}$ & 0.31737 \\
\hline \textbf{24} & $\Rz_{0,1}$ & 0.18838  & \textbf{55} & $\Rz_{0,1}$ & -1.31227  &  \textbf{86} & $\Rz_{0,1}$ & 1.60603  &  \textbf{117} & $\Rz_{0,1}$ & 0.02855  &  \textbf{148} & $\Rz_{0,1}$ & -0.00031 &  \textbf{179} & $\Rz_{0,1}$ & 0.26033 \\
\hline \textbf{25} & $\Rz_{1,2}$ & -0.20432  & \textbf{56} & $\Rz_{1,2}$ & 0.17508  &  \textbf{87} & $\Rz_{1,2}$ & 0.06147  &  \textbf{118} & $\Rz_{1,2}$ & 0.74683  &  \textbf{149} & $\Rz_{1,2}$ & 0.24798 &  \textbf{180} & $\Rz_{1,2}$ & 0.99595 \\
\hline \textbf{26} & $\Rz_{0,2}$ & 0.49788  & \textbf{57} & $\Rz_{0,2}$ & 0.26480  &  \textbf{88} & $\Rz_{0,2}$ & -0.01839  &  \textbf{119} & $\Rz_{0,2}$ & 0.19942  &  \textbf{150} & $\Rz_{0,2}$ & 0.33195 &  \textbf{181} & $\Rz_{0,2}$ & 0.62794 \\
\hline \textbf{27} & $\Rz_{2,3}$ & 0.58855  & \textbf{58} & $\Rz_{2,3}$ & 1.11408  &  \textbf{89} & $\Rz_{2,3}$ & -0.95135  &  \textbf{120} & $\Rz_{2,3}$ & 0.37599  &  \textbf{151} & $\Rz_{2,3}$ & 0.81622 &  \textbf{182} & $\Rz_{2,3}$ & -0.03170 \\
\hline \textbf{28} & $\Rz_{3,4}$ & 0.09731  & \textbf{59} & $\Rz_{3,4}$ & 0.13049  &  \textbf{90} & $\Rz_{3,4}$ & 1.39875  &  \textbf{121} & $\Rz_{3,4}$ & -0.13800  &  \textbf{152} & $\Rz_{3,4}$ & 0.41878 &  \textbf{183} & $\Rz_{3,4}$ & -0.14826 \\
\hline \textbf{29} & $\Rz_{4,5}$ & -1.04750  & \textbf{60} & $\Rz_{4,5}$ & 0.31341  &  \textbf{91} & $\Rz_{4,5}$ & 0.10627  &  \textbf{122} & $\Rz_{4,5}$ & 0.26683  &  \textbf{153} & $\Rz_{4,5}$ & 0.71386 &  \textbf{184} & $\Rz_{4,5}$ & 0.11074 \\
\hline \textbf{30} & $\Rz_{5,6}$ & 0.34316  & \textbf{61} & $\Rz_{5,6}$ & 1.52420  &  \textbf{92} & $\Rz_{5,6}$ & -1.00678  &  \textbf{123} & $\Rz_{5,6}$ & 0.63415  &  \textbf{154} & $\Rz_{5,6}$ & -0.83722 &  \textbf{185} & $\Rz_{5,6}$ & 0.25884 \\
\hline \textbf{31} & $\Rz_{6,4}$ & -0.12666  & \textbf{62} & $\Rz_{6,4}$ & 0.39354  &  \textbf{93} & $\Rz_{6,4}$ & -0.07047  &  \textbf{124} & $\Rz_{6,4}$ & 0.35838  &  \textbf{155} & $\Rz_{6,4}$ & 0.70935 &  \textbf{186} & $\Rz_{6,4}$ & 0.01146 \\
\hline
\end{tabular} %
}
\caption{The gates $\hat{O}$ and parameters $\theta$ which produce ansatz $\mathcal{A} \instate$, which is a result of Li simulation of the 7-qubit spin network for 700 iterations at $\Delta t=2.5\times 10^{-3}$. Note $Z_i = \exp(-i\theta \sigma^z_i /2)$, $Z_{i,j} = \exp(-i\theta \sigma^z_i \otimes \sigma^z_j /2)$ and similarly for $Y$ and $X$. These values are available in plaintext in the online repository~\cite{githubLink} (\href{https://github.com/QTechTheory/DissipativeRecompiler/blob/master/data/params0_final.txt}{direct link}).}
\label{tab:state_before_recomp}
\end{table}
}


\bibliographystyle{unsrt}

\end{document}